\documentclass[reprint,onecolumn,longbibliography,floats,aps,amsmath,amssymb,nofootinbib,prd,notitlepage]{revtex4-2}
\usepackage[a4paper,top=3cm,bottom=3cm,left=2cm,right=2cm,marginparwidth=1.75cm]{geometry}
\usepackage[english]{babel}
\usepackage[utf8]{inputenc}
\usepackage[T1]{fontenc}
\usepackage{amsmath}
\usepackage{amssymb}
\usepackage{lmodern}
\usepackage{color}
\usepackage{nccmath}
\usepackage{graphicx}
\usepackage{overpic}
\usepackage{wrapfig}
\usepackage{enumerate}
\usepackage{geometry}
\usepackage{braket}
\usepackage{orcidlink}
\usepackage{dsfont}
\usepackage{xcolor,soul,ulem}
\usepackage{hyperref}
\hypersetup{% Remove rectangles around links, keep only colored link.
    colorlinks,
    linkcolor={blue!90!black},
    citecolor={green!70!black},
    urlcolor={blue!90!black}
}

\usepackage{listingsutf8}
\usepackage{textgreek}
\definecolor{lightpurple}{rgb}{0.82, 0.75, 0.94}
\definecolor{TODOcolor}{HTML}{CCCC00}
\def\Listing#1{Listing~\ref{#1}}%
\def\Eq#1{(\ref{#1})}%

\definecolor{darkmagenta}{rgb}{0.55, 0.0, 0.55}

\definecolor{darkolivegreen}{rgb}{0.33, 0.42, 0.18}

\newcommand{\rd}{{\rm d}}
\newcommand{\re}{\mathbb{R}}
\newcommand{\Hil}{\mathcal{H}}
\newcommand{\lat}{\mathcal{L}}
\newcommand{\lPl}{\ell_{\rm Pl}}
\newcommand{\id}{\mathds{1}}

\begin{document}

\title{Integral Hilbert spaces and the dynamics of loop quantum cosmos}
\author{Janek Kozicki\,\orcidlink{0000-0002-8427-7263}}
	\email{jkozicki@pg.edu.pl}
	\affiliation{Faculty of Applied Physics and Mathematics, Gdańsk University of Technology, 80-233 Gdańsk, Poland}
	\affiliation{Advanced Materials Center, Gdańsk University of Technology, 80-233 Gdańsk, Poland}
\author{Tomasz Paw{\l}owski\,\orcidlink{0000-0002-5592-4804}}
	\email{tomasz.pawlowski@uwr.edu.pl}
	\affiliation{University of Wroc{\l}aw, Faculty of Physics and Astronomy, Institute for Theoretical Physics, pl. M. Borna 9, 50-204  Wroc{\l}aw, Poland}
\begin{abstract}
    Polymer quantization program applied in Loop Quantum Gravity/Cosmology leads to nonseparable Hilbert spaces. Commonly, one sidesteps this problem by singling out and working with a separable superselection sector. This is however often no longer accessible in more involved models beyond isotropic ones. In an alternative approach one builds a separable Hilbert space as an integral over all available sectors. Here we test the dynamics following from the latter on the example of a flat isotropic Universe admitting negative cosmological constant and a massless scalar field. There, numerical evolution of (initially) semiclassical states shows that there is no relevant difference in their long term semiclassicality properties in comparison to those in a single sector approach. Further, the older (problem-specific) numerical methods are compared against an application of more common and more efficient standard tools (\texttt{eigen} library).
\end{abstract}

\maketitle

\section{Introduction}
\label{sec:intro}

Among the attempts to unify the quantum and relativistic aspects of the reality description Loop Quantum Gravity (LQG) \cite{Rovelli:2004tv,Thiemann:2007pyv,Ashtekar:2004eh} is among the most recognized ones. Over the last couple of decades it reached a sufficient level of advancement to allow for a completion of a quantization program and at least in principle to control the dynamical evolution of quantum spacetimes \cite{Domagala:2010bm,Husain:2011tk,Giesel:2012rb}. However, high level of technical complication did not allow to extend the dynamical probing beyond unphysical simplest examples \cite{Zhang:2019dgi}. For that reason a lot of effort have been dedicated to its simplifications, starting from the so called reduced LQG \cite{Alesci:2013xd,*Alesci:2013xya,*Alesci:2014uha,Bilski:2016pib}, the (so called) midisuperspace approaches \cite{Bojowald:2004ag,*Bojowald:2004af,Gambini:2016czk,*Gambini:2020nsf,deBlas:2017goa} and the most radical (but also the most successful) Loop Quantum Cosmology (LQC) \cite{Ashtekar:2011ni, Bojowald:2008zzb}. They all share the same methodology, in particular the (following from the requirement of a strict background independence) choice of the nonstandard --the polymeric-- quantum representation. Their another defining feature is that the holonomies (parallel transports of connections) and fluxes (of the components densitized triads of orthonormal vectors) are used as fundamental candidates for quantization instead of metrics and its momentum and (due to the use of the polymeric representation) the latter may not even exist. As a consequence, one needs to rewrite basic observables and constraints of the theory/model in terms of these holonomies and fluxes (or their restrictions/analogs in simplified models) in a process known as Thiemann regularization.

An unfortunate consequence of using the polymeric representation is the nonseparability of the resulting Hilbert spaces, a feature persisting from full theory through all its simplifications (for an example of a simple harmonic oscillator see \cite{BarberoG:2013epp}). This property makes a completion of the quantization procedure and subsequent analysis of physical predictions challenging. In order to deal with this problem several ways have been explored by the community. One of them is the formalism of shadow states \cite{Ashtekar:2002sn}. Another, more intuitive, is the restriction of consideration to the superselection sectors -- mutually orthogonal subspaces closed with respect to the action of the evolution generators and the observables selected to describe particular physical system. In case of the full LQG these sectors are states (the so called spin networks) supported on disjoint graphs, whereas for its reductions/simplifications they depend on particular model. Each single sector is separable, thus upon restricting to it, one can treat them applying standard mathematical tools. This approach has been commonly used for example in studying isotropic universes in LQC. 

While this method works very well in isotropic sector, some problems appear already when moving to anisotropic models. An example is a simplest generalization -- a flat Bianchi I universe. For this model several prescriptions based on different choices for the Thiemann regularization have been considered. For most of them the convenient structure of superselection sectors persists \cite{Bojowald:2003md,*Bojowald:2003xf,Martin-Benito:2008dfr,*Martin-Benito:2009xaf,Diener:2017lde}, however for the construction that is considered to be the most physically consistent (in fact the only one of listed there, that admits a consistent and nontrivial infrared regulator  --fiducial cell-- removal limit) \cite{Ashtekar:2009vc} the superselection sector supports are dense in the domain of anisotropy configuration variables \cite{Martin-Benito:2011fdk} which makes the choice of a single sector unfeasible.

An alternative method of solving this problem has been proposed in \cite{BarberoG:2014ucb}. It uses the idea of integral Hilbert spaces used i.e. when quantizing a particle with periodic potential (see for example \cite{Reed:1979ne}). It utilizes the fact, that usually the spaces of superselection sectors admit a well defined measure. For example harmonic oscillator in \cite{BarberoG:2013epp} is a compact interval with a Lebesgue measure. This measure can then be used to define an integral over superselection sectors as well as the action of observables on it. Such integral Hilbert space will be separable, which again permits the use of the standard machinery to probe the physics of the system. However, its use a priori carries a risk of spoiling desirable properties featured by states in single superselection sectors, as even for simplest examples within LQC one sees slight differences in properties between the sectors, like slightly different behavior near the bounce or differences in discrete spectra of evolution operators (see for example \cite{Bentivegna2008} or \cite{MenaMarugan:2011me}). Thus, one cannot immediately dismiss the possibility that a state semiclassical before the bounce loses its semiclassicality (disperses) due to differences between sectors. The aim of presented work is to explicitly test the behavior of integral states in comparison with the single superselection sector ones on a relatively simple example. For that purpose we have selected a model of an isotropic FRLW universe admitting a massless scalar field and negative cosmological constant in LQC framework. Such model is quasi-cyclic -- the universe evolution features an infinite chain of bounces and recollapses, which allows one to probe the ``long term effects'' manifesting themselves after many cycles. On the other hand, the detailed analysis of the properties of different sectors --a desired basis for comparison-- is already available in the literature.

The article is composed as follows: in the following Section~\ref{sec:model} we introduce the example model and briefly recall the quantization procedure used in LQC in its context. Next, in  Section~\ref{sec:eigenspaces} we perform a detailed spectral analysis of the model's evolution operator, which is a central step in extracting the physical properties of the system. In particular, we discuss and compare the two distinct numerical approaches used to solve the eigenvalue problem. The results of this analysis are subsequently applied in Section~\ref{sec:dynamics}, where we construct a population of semiclassical quantum universes with Gaussian energy spectral profiles in both single sector and integral approach and perform a detailed comparison of both. Finally, the conclusions and closing remarks are presented in Section~\ref{sec:conclusions}.

\section{The example model}
\label{sec:model}

As mentioned above, the working example selected for our studies describes an isotropic Friedman-Lemaitre-Robertson-Walker (FRLW) universe admitting a negative cosmological constant and massless scalar field. Classically, in canonical description the state of the system is captured in two canonical pairs of global variables: scaled oriented volume $v$ (where the sign of $v$ corresponds to the orientation of the orthonormal triad encoding the spatial part of the metric), its canonical momentum $b$ proportional to the Hubble parameter, scalar field $\phi$ and its canonical momentum $p_{\phi}$. The relation between the volume $V$ of the selected comoving region of the Universe (the so called fiducial cell\footnote{In order to avoid infinities when integrating out Lagrangian/Hamiltonian densities over infinite homogeneous constant time slices one chooses a fiducial cell -- a ``finite box'' constant in comoving coordinates and restricts the integration to it. It thus plays a role analogous to an infrared regulator in quantum field theory.}), the Hubble parameter $H_r$ and the Poisson structure reads 
\begin{equation}\label{eq:dof}
  V = 2\pi\gamma G\hbar\sqrt{\Delta} |v| =: \alpha |v| \ , \quad 
  b = \gamma\sqrt{\Delta} H_r \ , \quad 
  \{b,v\} = 2/\hbar \ , \quad \{\phi,p_{\phi}\} = 1
\end{equation}
(with the remaining Poisson brackets vanishing), where $\gamma$ is the so called Barbero-Immirzi parameter \cite{BarberoG:1994eia} (dimensionless real constant of the order of $1$) and $\Delta=4\sqrt{3}\pi\gamma\approx5.17G\hbar$ is the so called LQC \textit{area gap} (see \cite{Ashtekar:2006wn} for details).

The quantization of this system in the framework of LQC has been described in details in \cite{Bentivegna2008}, though here we introduce slight modifications following \cite{Pawlowski2012} (which will be discussed further on). Since the system admits a constraint --the Hamiltonian one--
\begin{equation}\label{eq:NC-class}
  NC = p_{\phi}^2 - 3\pi\hbar^2 G b^2v^2 + \pi\gamma^2\Delta\hbar^2 G\Lambda v^2 = 0 \ ,
\end{equation}
where (unlike in \cite{Bentivegna2008}) the lapse function was set to $N=2V$ following \cite{Ashtekar:2007em}, the quantization procedure employs a Dirac program, splitting the process into two steps: kinematical level quantization (ignoring the constraint) and implementing the constraint.

The kinematical level quantization follows the procedure of \cite{Ashtekar:2003hd, Ashtekar:2006wn, Pawlowski2012}, in particular for the scalar matter the standard quantum mechanics techniques are used (Schr\"odinger representation), whereas the geometry is quantized via implementing the techniques of LQG. The resulting Hilbert space structure takes the form   
\begin{equation}
  \Hil = \Hil_{\rm gr}\otimes\Hil_{\phi} \ , \quad
  \Hil_{\rm gr} = L^2(\bar{\re},\rd\mu_{H}) \ , \quad 
  \Hil_{\phi} = L^2(\re,\rd\phi) \ ,
\end{equation}
where $\bar{\re}$ is the Bohr compactification of the real line and $\rd\mu_{H}$ is the Haar measure on it. A particularly convenient orthonormal basis on $\Hil_{\rm gr}$ can be formed out of eigenstates of the (oriented and rescaled -- see \eqref{eq:dof}) volume operator
\begin{equation}\label{eq:ip}
  \hat{v}\ket{v} = v\ket{v} \ , \quad
  \braket{v | v'} = \delta_{v,v'} \ ,
\end{equation}
where $\delta_{v,v'}$ is the Schr\"odinger delta. For $\Hil_{\phi}$ we will use the standard basis of (Dirac delta normalized) field generalized eigenstates $(\phi|$.

The constraints are implemented following \cite{Ashtekar:2006wn, Pawlowski2012}. First, the classical constraint (in the form corresponding to full general relativity in Ashtekar variables) is reexpressed in terms of holonomies along stright lines and the volume in the process of Thiemann regularization \cite{Thiemann:1997rt}. As this procedure is not unique, we implement the scheme known as the \textit{mainstream LQC} (specified in \cite{Ashtekar:2006wn})\footnote{For a systematic comparison of various schemes see for example \cite{Kowalczyk:2022ajp}.} The regularized constraint becomes then a function of $p_{\phi}$, $v$ and the $U(1)$ component of the holonomies $\mathcal{N} = \exp(ib/2)$.

Subsequently, $p_{\phi}$, $v$ and $\mathcal{N}$ are promoted to operators among which in selected basis $\hat{p}_{\phi}$ and $\hat{v}$ are multiplication ones, while $\hat{\mathcal{N}}$ becomes a shift operator
\begin{equation}
  \hat{\mathcal{N}}\ket{v} = \ket{v+1} \ .
\end{equation}
The resulting quantum constraint takes the form
\begin{equation}\label{eq:NC-quant}
  \widehat{NC} = \id \otimes \hat{p}_{\phi}^2 - \Theta_{\Lambda} \otimes \id \ , 
\end{equation}
and is well defined in the domain $\mathcal{D} = S^2(\bar{\re})\otimes\mathcal{S}(\re)$ where $S^2(\bar{\re})$ is the space of finite linear combinations sums of $\ket{v}, v\in\re$ and $\mathcal{S}(\re)$ is the Schwartz space. The operator $\Theta_{\Lambda}$ defined on $S^2(\bar{\re})$, known as the \textit{evolution operator} of LQC, is a 2nd order difference operator
\begin{equation}\label{eq:theta-Lambda}
  \Theta_{\Lambda} = \Theta_o - \pi G\gamma^2\Delta v^2\id \ , \quad
  \Theta_o = -\frac{3\pi G}{2} \left[\sqrt{|\hat{v}|}(\mathcal{N}^2-\mathcal{N}^{-2})\sqrt{|\hat{v}|}\right]^2 \ .
\end{equation}

Due to a discrete nature of the inner product \eqref{eq:ip} the space $\Hil_{\rm gr}$ (thus $\Hil$) is nonseparable. However, $\Theta_{\Lambda}$ naturally divides it onto mutually orthogonal subspaces $\Hil_{\epsilon}$ of states supported on the uniform lattices $\lat_{\epsilon}$
\begin{equation}
  \Hil_{\epsilon} := \Hil_{\rm gr} |_{\lat_{\epsilon}} \ , \quad 
  \lat_{\epsilon} := \epsilon + 4\mathbb{Z} \ , \ \epsilon\in[0,4) \ ,
\end{equation}
which are preserved by its action. These subspaces, known as superselection sectors, are separable.

Given that the most relevant geometry observables, like powers of volume, Hubble parameter, etc. preserve $\Hil_{\epsilon}$, one can restrict the studies to just one single sector, which is the route taken by the vast majority of works in LQC. 
Also, since no fermionic matter is considered here, one can further restrict the subspaces, exploiting the large symmetry of triad orientation (encoded in the sign of $v$) change, which distinguishes two sectors: symmetric and antisymmetric with respect to reflection in $v$. The symmetry of $\Theta_{\Lambda}$ and $\hat{V}$ with respect to $v\mapsto -v$ implies, that their action preserves those sectors. Following previous works, we also select the symmetric sector here. This complicates the original structure of the superselection sectors, as the reflection $v\mapsto -v$ transforms $\Hil_{\epsilon}$ into $\Hil_{4-\epsilon}$. We then distinguish three cases:
\begin{itemize}
  \item $\epsilon=0$: The structure of $\Theta_{\Lambda}$ itself (without any assumptions regarding the symmetry) naturally splits $\Hil_{\epsilon=0}$ onto three subspaces of states supported on $v=0$, $v\in 4\mathbb{Z}^+$ and $v\in 4\mathbb{Z}^-$ respectively, which are preserved by its action. As the state $\ket{v=0}$ completely decouples, it can be neglected when describing a large universe. As a consequence a symmetric state is uniquely characterized by its part supported on $\lat^+_0 = 4\mathbb{Z}^+$. 
  \item $\epsilon=2$: the reflection $v\mapsto -v$ preserves the subspace $\Hil_{\epsilon=2}$ and the requirement of symmetry (antisymmetry) imposes a set of nontrivial constraints on the states, namely $\forall n\in 2\mathbb{Z}^+, \forall \ket{\psi}\in\Hil_{\epsilon=2}\ \braket{n|\psi} = \pm \braket{-n|\psi}$ with a sign $+$ for symmetric and $-$ for the antisymmetric states respectively. For each of the choices the action of the operator $\Theta_o$ is the same as 
  \begin{equation}
    \tilde{\Theta}_o = -\frac{3\pi G}{2} 
      \left[ \mathcal{N}_4^+ \sqrt{|\hat{v}(\hat{v}+4)|}|\hat{v}+2| 
        - 2\hat{v}^2 + \mathcal{N}_4^- \sqrt{|\hat{v}(\hat{v}-4)|}|\hat{v}-2| \right]^2 \ ,
  \end{equation}
  where
  \begin{equation}
    \mathcal{N}_4^{\pm}\ket{v} 
    = \begin{cases} \ket{v+2} , & |v|>2 , \\ s\ket{v} , & |v|=2 , \end{cases}
  \end{equation}
  with $s=1$ for the symmetric and $s=-1$ for the antisymmetric sector respectively. Since the action of $\tilde{\Theta}_o$ splits $\Hil_{\epsilon=2}$ onto disjoint sub-sectors $\Hil_{\epsilon=2}^{\pm}$ of states supported on $\lat_{\epsilon=2}^{\pm} := \pm(2+4\mathbb{N})$, one can again restrict the studies to $\Hil_{\epsilon=2}^+$ while replacing $\Theta_o$ with $\tilde{\Theta}_o$ in \eqref{eq:theta-Lambda}
  \item $\epsilon\notin\{0,2\}$: for this generic case the reflection in $v$ mixes two distinct sectors $\Hil_{\epsilon}$ and $\Hil_{4-\epsilon}$, not generating any further constraints. By the symmetry of $\Theta_{\Lambda}$ only one of these sectors determines the entire pair. Consequently, imposing symmetry/antisymmetry effectively restricts the set of possible (generic) values of $\epsilon$ to either $(0,2)$ or $(2,4)$.
\end{itemize}

For the system studied (although with a different lapse $N=1$ and following from this choice slight differences in regularization prescription), the quantization program has been completed and the dynamics subsequently analyzed in \cite{Bentivegna2008}. The procedure consisted of the following steps
\begin{enumerate}
  \item The kernel of $\widehat{NC}$ has been found by group averaging procedure (see \cite{Ashtekar:1995zh}). The physical Hilbert space is spanned by the states represented (in volume and scalar field representation) by the wave functions\footnote{The full solution consists of two parts, corresponding to the positive and negative $\omega$, however following the procedure used for Klein-Gordon equation they are again considered superselection sectors and only the positive $\omega$ part is used.}
  \begin{equation}\label{eq:Psi-form}
      \Psi(v,\phi) = \sum_{k=0}^{\infty} \tilde{\Psi}_n e_n(v) e^{i\omega_n\phi} \ ,
  \end{equation}
  where $\omega_n^2$ were the elements of the spectrum of $\Theta_{\lambda}$ (which in this case is purely discrete and nondegenerate), $e_n$ are the normalized eigenfunctions corresponding to the spectrum elements and $\tilde{\Psi}\in S^2(\mathbb{N})$ is a discrete spectral profile (a square summable sequence). 
  \item In order to define a nontrivial dynamics the notion of partial observables \cite{Rovelli:2001bz,Dittrich:2004cb} was used. Specifically, the field $\phi$ has been considered as an evolution parameter and subsequently the set of observables corresponding to values of geometry observables at a given value $\phi$ has been constructed. The strict procedure of constructing such family is described for example in \cite{Kaminski:2009qb}, although its result is mathematically equivalent to a (quantum level) deparametrization with respect to $\phi$, where the constraint \eqref{eq:NC-quant} is interpreted as a Klein-Gordon type evolution equation. Upon that, the system can be considered as an evolution of a geometry state $\psi\in\Hil_{\rm gr}$ with respect to a matter clock $\phi$ and generated by the operator $\sqrt{|\Theta_{\lambda}|}$. In that picture any kinematical geometry observable $\hat{O}$ is automatically lifted to a (partial) physical observable $\hat{O}_{\phi}$ measuring a particular quantity at a time $\phi$. 
  \item Once both physical states and a sufficiently large set of physically interesting observables were available, a set of semiclassical states (in particular in \cite{Bentivegna2008} appropriate cutoffs of Gaussian spectral profiles to ${\rm Sp}(\Theta_{\Lambda})$) has been selected and the quantum trajectories (expectation values of the observables as functions of $\phi$) were evaluated.
\end{enumerate}

Let us now consider an alternative to choosing a single superselection sector $\Hil_{\omega}$. Since the sectors are parametrized by values of $\epsilon \in [0,4)$, which in turn has a natural Lebesgue measure $\rd\epsilon$, one can easily construct an integral Hilbert space \cite{BarberoG:2014ucb}
\begin{equation}
  \Hil_{\rm int} := \frac{1}{\bar{V}} \int_{0}^{4} \Hil_{\epsilon} \rd\epsilon \ , \quad 
  \braket{\Psi|\Phi}_{\Hil_{\rm int}} 
  = \frac{1}{\bar{V}} \int_{0}^{4} \braket{\Psi_{\epsilon}|\Phi_{\epsilon}}_{\Hil_{\epsilon}}{\rd\epsilon} , \quad
  \Psi_{\epsilon} = \Psi|_{\lat_{\epsilon}} , 
  \Phi_{\epsilon} = \Phi|_{\lat_{\epsilon}} \ ,
\end{equation}
where the factor $\bar{V} := \int_0^4\rd\epsilon = 4$ is the volume of the set of sectors.
Similarly, an action of any partial observable $\hat{O}_{\phi}$ can be defined through its restrictions to $\Hil_{\epsilon}$. Consequently, its expectation value can be evaluated as
\begin{equation}\label{eq:int-O}
  \braket{\Psi|\hat{O}_{\phi}|\Psi}_{\Hil_{\rm int}} = 
  \frac{1}{4} \int_{0}^{4} \braket{\Psi_{\epsilon}|\hat{O}_{\phi}|\Psi_{\epsilon}}_{\Hil_{\epsilon}} \rd\epsilon \ .
\end{equation}
Note, that a different choice of a Lebesgue measure $\rd\epsilon\mapsto f(\epsilon)\rd\epsilon$ will lead to a unitarily equivalent space.

The above construction allows to perform a completion of the quantization procedure by first performing the points 1-3 for a sufficiently large set of superselection sectors and subsequently integrate out the results (the quantum trajectories) via \eqref{eq:int-O}. We will perform this step in Section~\ref{sec:dynamics}. Before that however, let us examine in detail the spectral properties of $\Theta_{\Lambda}$.

\section{Eigenspaces of the evolution operator}
\label{sec:eigenspaces}

Here, we start with a precise mathematical specification of the eigenvalue problem. Next we discuss and compare two distinct numerical methods used to solve it.

\subsection{The eigenvalue problem for the evolution operator}

To start with, let us note, that due to non-negativity of the operator $\Theta_o$ \cite{Ashtekar:2007em} the operator $\Theta_{\Lambda}$ is also nonnegative. As a consequence the generalized eigenvalue problem for it can be (in $v$ representation) formulated in the following way
\begin{equation}\label{eq:eigensystem}
  \Theta_{\Lambda} \psi(v) = \omega^2 \psi(v). 
\end{equation}
which yields a 2nd order difference equation
\begin{equation}
\label{eq:2ndOrder}
  f_-(v)\psi(v-4) + (\omega^2 + \pi G\gamma^2\Delta\Lambda v^2 - f_o(v) )\psi(v) + f_+(v)\psi(v+4) = 0 \ ,
\end{equation}
where
\begin{subequations}\label{eq:2ndOrder-coeff}\begin{align}
  f_{\pm}(v) &= (3 \pi G / 4)\sqrt{|v(v\pm 4)|}|v\pm 2| \ , \label{eq:f_plus_minus} \\
  f_{o}(v) &=(3 \pi G / 2) v^2 \ . \label{eq:f_zero}
\end{align}\end{subequations}
This equation can easily be solved iteratively, however, once we restrict ourselves to symmetric states only, the structure of free degrees of freedom depends on the particular sector. Specifically:
\begin{itemize}
  \item For \textit{exceptional sectors} $\epsilon\in\{0,2\}$ The generalized eigenspaces are 1-dimensional and entire eigenfunction is determined by the value $\psi(v=4)$ and $\psi(v=2)$ respectively.
  \item For \textit{generic sectors} we need a pair of values at two consecutive points, say $\psi(\epsilon)$ and $\psi(\epsilon+4)$ to uniquely determine the solution.
\end{itemize}

Since the coefficients of \eqref{eq:eigensystem} are real, it is enough to solve the eigenvalue problem for real and imaginary part separately, thus one can assume without the loss of generality, that the generalized eigenfunctions are real. To solve the specified problem, one could in principle transform it to the representation in momentum $b$ of $v$ as it was done in \cite{Pawlowski2012} for $\Lambda>0$. However, applying this technique for generic lattices becomes mathematically involved, as the operator $|\hat{v}|$ does not transform into a local operator in $b$ representation. In the case $\epsilon=0$ one exploits the decoupling of $v>0$ and $v<0$ sectors and the fact that for each of these sectors separately this operator does transform into a local one. In our case however, where we need to probe entire range of $\epsilon$ such structure is no longer available. Therefore, we will resort to solving the equation \eqref{eq:eigensystem} directly in $v$ representation. This will be done by using two techniques: the ``shots'' method originally used in \cite{Bentivegna2008} and a new one employing the \texttt{eigen} library. Let us start with the former.

\subsection{Numerical solutions: shots method}
\label{sec:shots}

\begin{figure}[p]
  \includegraphics[width=0.8\textwidth]{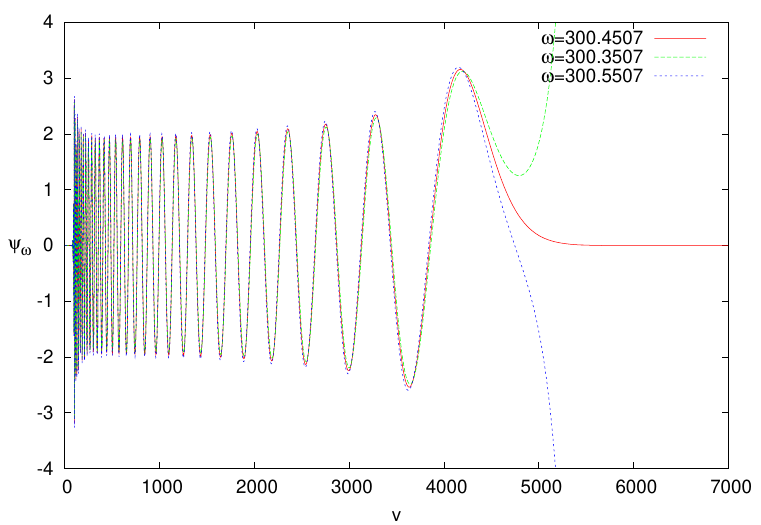}
  \caption{An illustration of the shots method functionality for the exceptional sector $\epsilon=0$ on the example of the model considered in this work quantized with older APS prescription. The identified eigenfunction is bracketed by two ones with growing tails. Source: \cite{Bentivegna2008}}
  \label{fig:shots-example}
\end{figure}

The shots method is based on a key observation regarding the general solutions to the eigenvalue problem discussed already in \cite{Bentivegna2008}. For $|v|$ exceeding certain ($\omega$ dependent) value $v_r(\omega)$ approximately corresponding to the recollapse point of the classical trajectory of universe with $p_{\phi}=\hbar\omega$, one observes an exponential behavior (either growth or decay). With the inner product given by \eqref{eq:ip} only the (nongeneric) eigenfunctions decaying in specified zones contribute to the decomposition of unity and the values $\omega^2$ corresponding to them belong to the spectrum of $\Theta_{\Lambda}$. Furthermore, the solutions to \eqref{eq:eigensystem} are continuous in initial data and those with decaying tails are explicitly normalizable. In consequence, the initial data corresponding to the decaying solutions (forming a set of isolated points in the space of initial data) can be numerically identified as the ones, for which the (generically growing) exponential tails change the sign (see Fig.~\ref{fig:shots-example}). 

Because of the different dimensionality of the space of (free) initial data for the exceptional $(\epsilon\in\{0,2\})$ and generic $(\epsilon\notin\{0,2\})$ sectors, for each of them the algorithm of finding the normalizable eigenfunctions is slightly different.
\begin{itemize}
  \item In exceptional sectors for given $\omega$ the value of $\psi$ at either $v=4$ (for $\epsilon=0$) or $v=2$ (for $\epsilon=2$) determines the whole eigenfunction $\psi_{\omega}$ uniquely and scaling the initial data corresponds to global scaling of the whole $\psi_{\omega}$, thus it can be set to $1$ without the loss of generality. Identification of the normalizable eigenfunctions is performed by identifying $\omega$'s for which the growing exponential tail at $v>v_r$ changes sign. This is performed by probing said sign on a uniform lattice in $\omega$ (with separation $\Delta\omega = 0.1G^{1/2}$) and subsequent bissection.
  \item In generic sectors for given $\omega$ the space of initial data is $2$-dimensional: $\psi_{\omega}(v_i), \psi_{\omega}(v_i+4)$ (where the choice of $v_i$ will be discussed later), however a simultaneous scaling of both values again corresponds to a global rescaling of the entire eigenfunction. Thus, without the loss of generality one can parametrize the data as $\psi_{\omega}(v_i) = \cos(\varphi), \psi_{\omega}(v_i+4) = \sin(\varphi)$ where $\varphi\in [0,2\pi)$. Finding the normalizable eigenfunctions boils down to an identification of (isolated) pairs $(\omega,\varphi)$ for which the tails at $|v|>v_r$ decay for both positive and negative $v$. This is done via an extension of the algorithm used for exceptional lattices:
  \begin{itemize}
    \item First, for each $\omega$ the parameter $\varphi$ corresponding to a solution with a tail decaying at one (say $v>0$) side is identified via a direct analog of the algorithm for exceptional sector: $\varphi$ is scanned in the uniform lattice with separation $\Delta\varphi = 0.05\pi$ and the values of $\varphi$ for which the tail flips the sign are found via a bissection.
    \item Next, the results of the first step are used in the algorithm for exceptional sector (though with the step separation $\Delta\omega$ changed to $\max(0.005,0.5*(1-|\epsilon-1|))G^{1/2}$) to identify the values $(\omega, \varphi)$ where the sign of the tail on the opposite side flips.
  \end{itemize}
\end{itemize}
Since for $|v|\gg v_r$ the problem is numerically unstable, even upon identification of the data $\omega$ or $(\omega,\varphi)$ with relative precision of $10^{-15}$ the tails eventually grow due to accumulation of the numerical errors. Therefore, the solutions $\psi_{\omega}$ need to be further regularized. This is done by finding the first (looking from the edge of the domain) minimum of $|\psi_{\omega}|$ on each side and removing the (growing) tails. In \cite{Bentivegna2008} this was done by setting $\psi_{\omega}(v)$ to zero past these points, however this estimate has proven too crude for the current work. Therefore, the tails were extrapolated via assuming the pure exponential decay, of which exponent has been interpolated from values  of $\psi_{\omega}$ in an interval of $|v|$ slightly before the identified minima of $\psi_{\omega}$. In principle, the precision of evaluating the tails could be improved by substituting them with the tails of the eigenfunctions corresponding to the Wheeler-DeWitt analog of the studied system (see \cite{Bentivegna2008}, sec.~III.B). However, these eigenfunctions are expressed via Bessel functions of the third kind of imaginary order, numerical evaluation of which with sufficient precision is again a bit involved.

% properties
After the regularization procedure, the eigenfunctions were normalized (with required norm evaluated via a direct summation). With exception of the few lowest $\omega$'s (for which the eigenfunctions were peaked around $v=0$) they share the following properties already observed for a similar system in \cite{Bentivegna2008}.
\begin{itemize}
  \item One can distinguish three zones separated by certain $\omega$-dependent values $v_b, v_r$: $(i)$ interior \emph{sub-bounce} one $|v|<v_b(\omega)$ where the behavior of $\psi_{\omega}$ is quasi-exponential, $(ii)$ \emph{classical} $v_b(\omega) < |v| < v_r(\omega)$, where $\psi_{\omega}$ oscillates, and $(iii)$ \emph{past-recollapse} one where it decays (quasi)exponentially. 
  \item For generic sectors the eigenfunctions are highly imbalanced, with one orientation (sign of $v$) strongly suppressed (by a factor approximately exponential in $\omega$). The orientation of the suppressed side alternates for consecutive eigenfunctions.
\end{itemize}
The values $v_b(\omega)$, $v_r(\omega)$ are the points, where the transition matrix of the initial value problem for the generalized eigenfunction cast as a $1$st order difference equation (see Appendix~\ref{app:stability}) changes the structure: its eigenvalues change from real ($|v|<v_b$ or $|v|>v_r$) to complex ($v_b<|v|<v_r$). Their values can be evaluated explicitly (see eq.~\eqref{eq:vbr-app}), though they can be approximated with a decent accuracy via the position of the bounce of the LQC flat universe with $\langle \hat{p}_{\phi} \rangle = \hbar\omega$ and $\Lambda=0$ (approximated by the $0$th order effective dynamics \cite{Singh:2005xg}) and the recollapse position of the classical universe with $p_{\phi} = \hbar\omega$ and negative cosmological constant $\Lambda$ respectively \begin{equation}\label{eq:vbr-approx}
  v_b(\omega) \approx \tilde{v}_b(\omega) 
  := \frac{\omega}{\sqrt{3\pi G}} \ ,
  \qquad 
  v_r(\omega) \approx \tilde{v}_r(\omega) 
  := \frac{\omega}{\sqrt{-\pi G\gamma^2\Delta\Lambda}} \ . 
\end{equation}

As in the sub-bounce region the problem is unstable and only the solutions growing (in that zone) for consecutive iterative steps were determined with sufficient reliability, for generic sectors certain care had to be taken in regards of specifying the position $v_i$ of the initial data: because of the high inbalance and the instability in the sub-bounce region the data needs to be specified within classically allowed or past-recollapse region and the procedure of finding the normalizable solutions described above will work correctly only for the eigenfunctions suppressed on the side (orientation of $v$) where the data is specified. Therefore, the process of scanning for the solutions has been performed in two parallel flows, with the position of the initial data set a bit beyond recollapse in $v>0$ and $v<0$ respectively. In actual simulations that value has been set as the lattice $\lat_{\epsilon}$ point closest to $|v|_{\rm id} = \max(\tilde{v}_r+300,500)$, where $\tilde{v}_r$ is given by \eqref{eq:vbr-approx}.

\subsection{Numerical solutions: application of the \texttt{eigen} library}
\label{sec:eigen}

The eigenfunctions of \Eq{eq:eigensystem} were found using the \texttt{eigen}
library~\cite{Eigen2020} class \texttt{Eigen::SelfAdjointEigenSolver} and its
built in method \texttt{computeFromTridiagonal}. The tridiagonal matrix
$\Theta_{\Lambda}$ in the following equation (rearranged from \Eq{eq:2ndOrder}):
\begin{equation}
\label{eq:Theta_psi}
-[\Theta_{\Lambda} \psi](v)=f_{-}(v)\psi(v-4)+\left(\pi G\gamma^2\Delta\Lambda v^2-f_o(v)\right)\psi(v)+f_{+}(v)\psi(v+4),
\end{equation}
\noindent
is built explicitly in the function
\texttt{getMatrixTheta()} shown on \Listing{listing1}. Please note that LHS of \Eq{eq:Theta_psi} is negative, and in the \texttt{getMatrixTheta()} function it is build as positive, hence all terms in the listing are with negative sign. In this listing the
function \texttt{FRW\_func\_f\_PlusMinus} corresponds to  \Eq{eq:f_plus_minus} and \texttt{FRW\_func\_f\_Zero\_and\_Lambda}
corresponds to \Eq{eq:f_zero} and the term $\pi G \gamma^2 \Delta \Lambda v^2$.

An important point is that, as can be seen in \Listing{listing1}, the matrix representation of operator 
$\Theta_{\Lambda}$ is cut at the matrix boundary, which corresponds to a potential barrier at the boundary which is not the case in the real physical system. To compensate for this numerical deficiency of the numerical representation of operator $\Theta_{\Lambda}$ the matrix is made significantly bigger and only the lowest $10\%$ of the found eigenvalues and eigenfunctions are used as the solutions. These lowest eigen solutions are not affected by the potential barrier at the boundary, because the boundary is too far away from them.

Next, in the
\Listing{listing2} the eigen system \Eq{eq:eigensystem} of this matrix is
solved using the \texttt{computeFromTridiagonal} method from the \texttt{eigen}
library~\cite{Eigen2020}.

\begin{figure*}[htbp]
\begin{minipage}{\textwidth}
\begin{lstlisting}[numbers=left,firstline=1,lastline=19,label=listing1,
caption={Construction of the tridiagonal $\Theta_o$ matrix in order to find its eigenvalues and eigenvectors.},
language=c]
MatrixXr State::getMatrixTheta()
{
	int N = config.points;
	MatrixXr theta = MatrixXr::Zero(N, N);
	for(int i = 0 ; i < N ; i++ ) {
		Real v        = i*4 + config.epsilonFRW - 2*N;  // calculate v with ε shift
		Real f_Plus   = FRW_func_f_PlusMinus(v, true);  // Eq.16a with plus
		Real f_Minus  = FRW_func_f_PlusMinus(v, false); // Eq.16a with minus
		Real f_Zero_L = FRW_func_f_Zero_and_Lambda(v);  // Eq.16b - πGγ²∆Λv²
		theta(i,i)    = f_Zero_L;
		if(i > 0 ) {
			theta(i, i - 1) = -f_Minus;
		}
		if(i < N - 1 ) {
			theta(i, i + 1) = -f_Plus;
		}
	}
	return theta;
}
\end{lstlisting}
\end{minipage}%\\[-8mm]
\end{figure*}

\begin{figure*}[htbp]
\begin{minipage}{\textwidth}
\begin{lstlisting}[numbers=left,firstline=1,lastline=12,label=listing2,
caption={Solving the eigen system of $\Theta_o$ matrix.},
language=c]
void State::calc_computeFromTridiagonal()
{
	MatrixXr theta       = getMatrixTheta();
	VectorXr diagonal    = theta.diagonal(0);
	VectorXr subDiagonal = theta.diagonal(-1);

	Eigen::SelfAdjointEigenSolver<MatrixXr> solver;
	solver.computeFromTridiagonal(diagonal, subDiagonal);

	saveEigenvalues(solver.eigenvalues());
	saveEigenvectors(solver.eigenvectors());
}
\end{lstlisting}
\end{minipage}%\\[-8mm]
\end{figure*}

\subsection{Comparison of numerical costs}

The shots method described in Sec.~\ref{sec:shots} has been constructed specifically for the identification of discrete spectra of operators in LQC -- it utilizes the specifics of the eigenvalue problem for these operators and is not affected by the noncompactness of the domain of $v$. Consequently, when using it, it is enough to scan only the specific range of the potential eigenvalues in which we intend to identify the spectrum for further calculations (i.e. for the construction of a particular set of semiclassical states). In that sense it is optimized for the task to which it is applied and significantly more efficient than direct applications of standard methods. For example, in the computations of the most demanding cases probed in presented work ($\Lambda=-0.05$, states peaked about $\omega_o=4000G^{1/2}$) it was sufficient to find only about $3400$ lowest eigenvalues (and eigenstates corresponding to them), which to solve the eigenvalue problem for $64$ superselection sectors took in total about $8$ days on the workstation equipped with an AMD Ryzen 9 5950X processor (16 cores) and $64$GB of RAM (utilizing 32 parallel processes) utilizing below $30\%$ of the available RAM.

In the method directly applying the \texttt{eigen} library~\cite{Eigen2020} the standard generic method for solving eigenvalues problems is used. As that method requires representing the operator by a finite matrix, a truncation of the original $\hat{\Theta}_{\Lambda}$ is necessary. Physically, such truncation corresponds to an infinite potential barrier at the boundary of the domain. Consequently, in order to keep the errors due to truncation sufficiently low, it is necessary to define the (truncated) operator in sufficiently large domain of $v$ -- significantly larger than when using the previous method. In consequence, the library solves the eigenvalue problem for much larger domain of potential spectrum elements, of which only the lowest portion is evaluated with sufficient accuracy for the goals of our studies. In actual calculations only $10\%$ of the range of eigenvalues was used. As a consequence, for the most demanding case (specified in the paragraph above) instead of just 3400 eigenvalues (and eigenfunctions) one needed to calculate about 34000 of them. The problem was solved for each $\epsilon$ (of 64 of them in total) in separate thread, each thread using 45 GB of RAM, thus on a server with 1 TB of RAM, 22 of them were running in parallel, taking about 5 days (on AMD EPYC 7702P processor with 128 threads). Consequently, in order to calculate the spectrum of $\hat{\Theta}_{\Lambda}$ for all 64 superselection sectors the calculations were performed in three batches, taking in total 15 days.

\section{The dynamics of the example Universe}
\label{sec:dynamics}

The energy spectrum and the basis formed of energy eigenstates constructed in the previous section can now be used to assemble physical states of desired properties. Here we are interested in large (within the limits of spectrum possible to probe by our methods) semiclassical universe, thus we will focus our studies on the ones most commonly used in the field -- those of Gaussian spectral profile. In this section we will construct a population of such states for both single superselection sectors and integral approach, subsequently probing and comparing their dynamical properties, especially the process of decoherence and semiclassicality loss.
Let us start with the construction of the states and the particular quantities used for their description.

\subsection{Semiclassical states and their dynamics}

In Section~\ref{sec:eigenspaces} the eigenfunctions forming an energy basis were found (as functions of $v$) for each superselection sector $\Hil_{\epsilon}$ separately and labeled by natural numbers due to discreteness of the spectrum. In order to adapt the notation for the (future) integration over the superselection sectors here we will denote each basis element $e_j$ within a sector corresponding to a given $\epsilon$ as $e_{\epsilon,j}(v)$ and the eigenvalue of the operator 
$\sqrt{\Theta_\Lambda}$ (corresponding to it) as $\omega_{\epsilon,j}$.

For chosen $\epsilon$ each physical state $\ket{\Psi}_{\epsilon}$ -- a solution to the constraint \eqref{eq:NC-quant} found by group averaging (see Sec.~\ref{sec:model}) can be represented by its spectral profile $\tilde{\Psi}_{\epsilon}\in\Sigma^2(\mathbb{N})$ so that its wave function in $v$ representation is given by \eqref{eq:Psi-form}, which in proposed notation takes the form
\begin{equation}
\label{eq:timeEvol}
\Psi_\epsilon(v,\phi)=\sum_j \tilde{\Psi}_{\epsilon,j} e_{\epsilon,j}(v) e^{i \omega_{\epsilon,j} \phi} \ .
\end{equation}
As stated earlier, we focus our investigation on a specific class of states with Gaussian profile (peaked about $\omega_o$ with a variance in $\omega$ equal to $\sigma_{\omega}/\sqrt{2}$), specifically we take the spectral profile to be of the form 
\begin{equation}
\label{eq:projection}
\tilde{\Psi}_{\epsilon,j} 
= \frac{1}{\sigma_\omega \sqrt{2 \pi}}e^{-\frac{1}{2}\left(\frac{\omega-\omega_o}{\sigma_\omega}\right)^2}.
\end{equation}
The notion of dynamics is introduced through the family of partial observables $\hat{O}_{\phi}$ parametrized by the value of the scalar field (see again Sec.~\ref{sec:model}). Since (for the model studied) mathematically such picture is equivalent to reinterpreting the constraint \eqref{eq:NC-quant} as a free Klein-Gordon-like evolution equation of the physical state represented by a constant $\phi$ slice of $\ket{\Psi}$, we determine the action of $\hat{O}_{\phi}$ as action of the adequate observable $\hat{O}$ (acting on some domain in $\Hil_{\rm gr}$) on particular slice $\ket{\Psi(\cdot,\phi)}$. Here, we focus our attention on the volume $\hat{V}_{\phi}$ of the Universe (region) at given $\phi$. For given superselection sector $\epsilon$ its expectation value and variation is calculated as follows
\begin{subequations}\begin{align}
  \left<V\right>_\epsilon 
  &= \alpha \sum_v \Psi_\epsilon(v,\phi)^{*}\, |v| \, \Psi_\epsilon(v,\phi) \ , 
  \label{eq:avg_v_eps} \\
  \left<V^2\right>_\epsilon 
  &= \alpha^2 \sum_v \psi_\epsilon(v,\phi)^{*}\, v^2 \, \psi_\epsilon(v,\phi) \ , 
  \label{eq:avg_v2_eps} \\
  \Delta V_\epsilon 
  &= \sqrt{\left<V^2\right>_\epsilon - \left<V\right>^2_\epsilon} \ ,
  \label{eq:stdDev_v_eps}
\end{align}\end{subequations}
where the constant $\alpha$ is defined in \eqref{eq:dof}. 

Having at out disposal $\braket{V}$ and $\Delta V$ for each particular superselection sector we can evaluate them on the superselection sector integral states via \eqref{eq:int-O}. Specifically
\begin{subequations}\label{eq:IntAll}\begin{align}
  \left<V\right> 
  &= \frac{1}{4} \int_0^4 \left<V\right>_\epsilon\, \rd\epsilon \ ,
  \label{eq:vIntRomberg} \\
  \left<V^2\right> 
  &= \frac{1}{4} \int_0^4 \left<V^2\right>_\epsilon\, \rd\epsilon \ ,
  \label{eq:v2IntRomberg} \\
  \Delta V 
  &= \sqrt{\left<V^2\right> - \left<V\right>^2} \ .
  \label{eq:stdDev_v}
\end{align}\end{subequations}

In actual simulations the integration was performed via Romberg method on a grid of $64$ values of $\epsilon$ uniformly distributed over an interval $[0,4)$. The calculations were performed for a set of values of cosmological constant $\Lambda\in\{-0.2,-0.1,-0.05,-0.01\}\ell_{\rm Pl}^{-2}$ for semiclassical states peaked about $\omega\in\{10^3,4\cdot 10^3\}G^{1/2}$ with variance $\Delta\omega\in\{50/\sqrt{2},200/\sqrt{2}\}G^{1/2}$.
Their results are used in the next subsection to compare the properties of single sector versus integral states.

\subsection{The results}

Let us recall, that the main question motivating presented work is whether in the framework used there exist relevant differences in the dynamics of a (semiclassical) quantum Universe between superselection sectors, in particular whether such differences could lead to an alteration of the dynamics in the integral approach. In order to address this question we performed a systematic comparison of the expectation values and variances of the volume $\hat{V}_{\phi}$ as functions of $\phi$. Upon selecting a particular Gaussian profile (for chosen $\omega_o$ and $\sigma_{\omega}$) a family of single sector states was defined by the projection \eqref{eq:projection} of this profile to the appropriate spectra of $\hat{\Theta}_{\Lambda}$. Subsequently, expectation values of volume $\langle V \rangle_\epsilon$ and their variances $\Delta V_\epsilon$ were evaluated (as functions of both $\phi$ and $\epsilon$) via \eqref{eq:avg_v_eps} and \eqref{eq:stdDev_v_eps} and finally integrated via \eqref{eq:IntAll} to give the integral trajectories and variances $\langle V \rangle$ and $\Delta V$ (as functions of $\phi$). To further verify the robustness of the results the calculations above have been performed with use of the  basis evaluated via two distinct methods: shots one (Sec.~\ref{sec:shots} and using \texttt{eigen} library (Sec.~\ref{sec:eigen}). 

\begin{figure*}[p]
    \begin{overpic}[width=\textwidth]{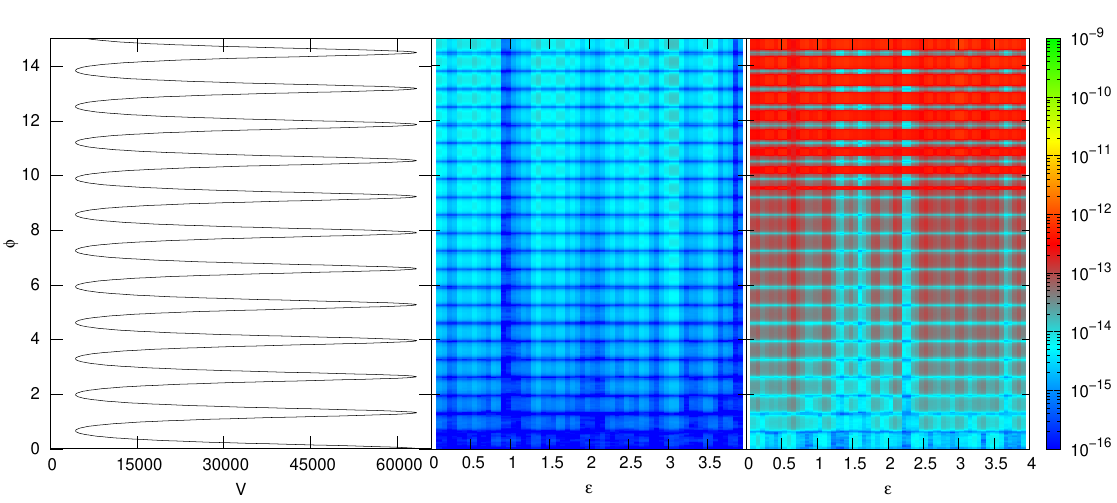}
        \put(21,1.0){\makebox(0,0)[c]{\color{white}\rule{1.5em}{2.5ex}}}
        \put(21,0.5){$V$}
        \put(53,1.0){\makebox(0,0)[c]{\color{white}\rule{1.5em}{2.5ex}}}
        \put(53,0.5){$\epsilon$}
        \put(79,1.0){\makebox(0,0)[c]{\color{white}\rule{1.5em}{2.5ex}}}
        \put(79,0.5){$\epsilon$}
        \put(0.5,22.5){\makebox(0,0)[c]{\color{white}\rule{2.0em}{3.5ex}}}
        \put(0.5,22.5){$\phi$}
        \put(4,42.5){(a)}
        \put(38,42.5){(b)}
        \put(67,42.5){(c)}
	 \put(-1,44){(A)}
    \end{overpic}
    \begin{overpic}[width=\textwidth]{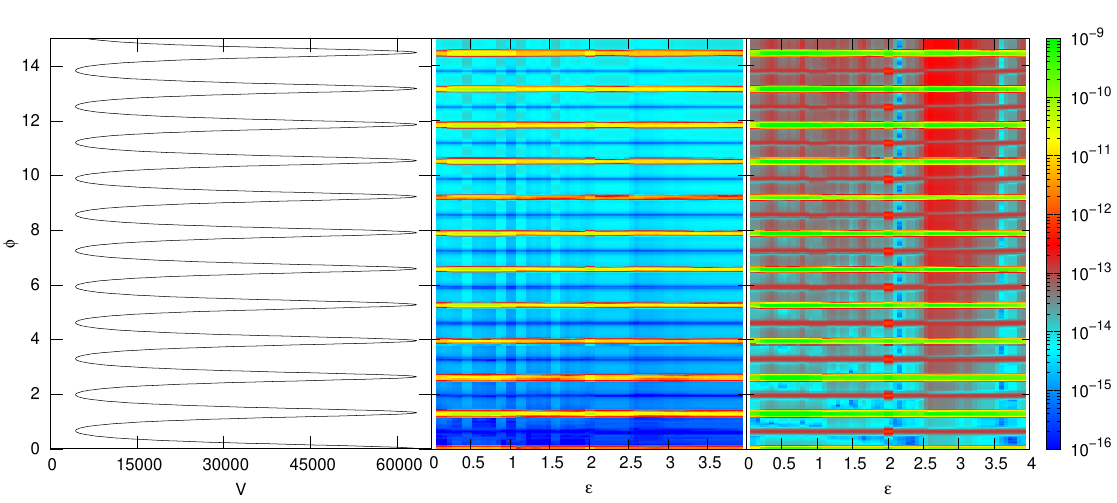}
        \put(21,1.0){\makebox(0,0)[c]{\color{white}\rule{1.5em}{2.5ex}}}
        \put(21,0.5){$V$}
        \put(53,1.0){\makebox(0,0)[c]{\color{white}\rule{1.5em}{2.5ex}}}
        \put(53,0.5){$\epsilon$}
        \put(79,1.0){\makebox(0,0)[c]{\color{white}\rule{1.5em}{2.5ex}}}
        \put(79,0.5){$\epsilon$}
        \put(0.5,22.5){\makebox(0,0)[c]{\color{white}\rule{2.0em}{3.5ex}}}
        \put(0.5,22.5){$\phi$}
        \put(4,42.5){(a)}
        \put(38,42.5){(b)}
        \put(67,42.5){(c)}
	 \put(-1,44){(B)}
    \end{overpic}
    \caption{A map of relative differences $\delta V_r$ (middle column) and $\delta\sigma_r$ (right column) \eqref{eq:RelDiffDef} in $\langle V\rangle$ and $\Delta V$ between single sector and integral Gaussian states peaked about $\omega_o=4000G^{1/2}$ (with $\sigma_{\omega}=200G^{1/2}$) for $\Lambda=-0.05$ constructed form basis generated by \texttt{eigen} (upper row) and shots method (lower row). Left column shows the trajectory $\langle V\rangle(\phi)$ to help associate particular value of $\phi$ with the stage of the evolution.}
    \label{fig:bounceMap2}
\end{figure*}

\begin{figure*}[p]
    \begin{overpic}[width=0.83\textwidth]{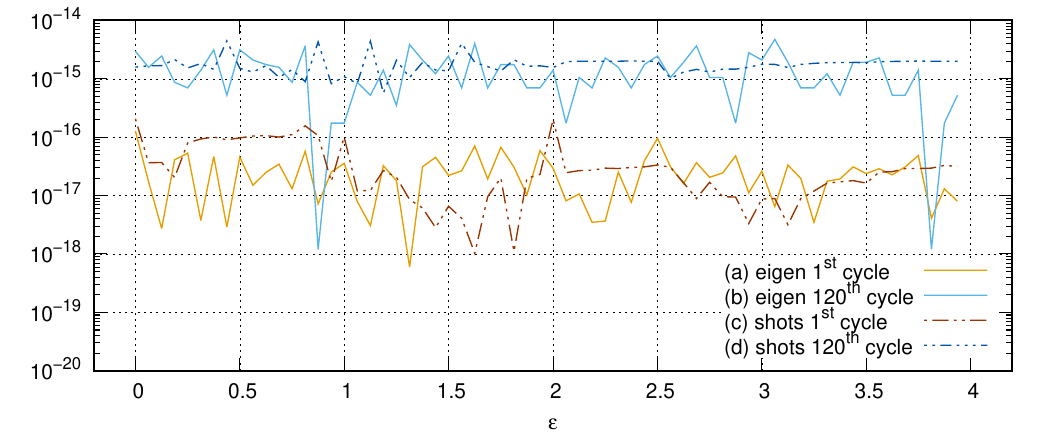}
         \put(52.6,1.6){\makebox(0,0)[c]{\color{white}\rule{1.5em}{2.5ex}}}
         \put(52,1){$\epsilon$}
         \put(-3,22.2){$|\delta V_r|$}
	 \put(-1,42){(A)}
    \end{overpic}
    \begin{overpic}[width=0.83\textwidth]{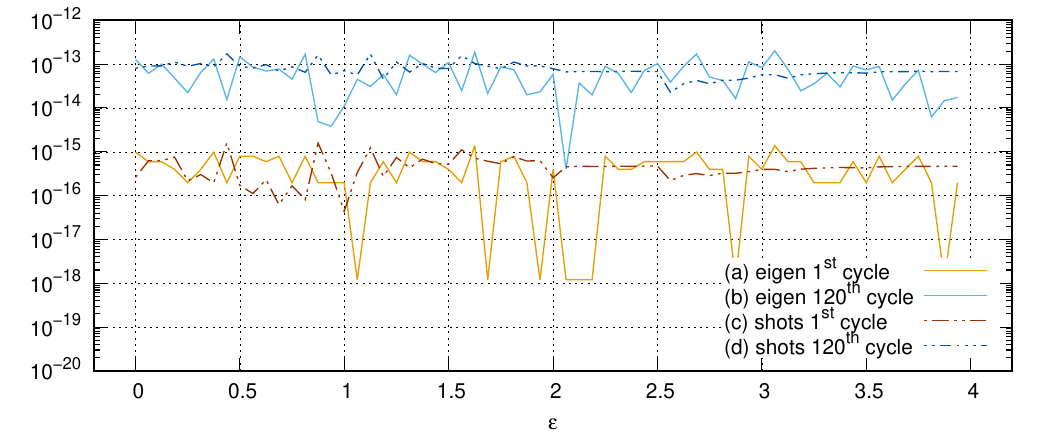}
         \put(52.6,1.6){\makebox(0,0)[c]{\color{white}\rule{1.5em}{2.5ex}}}
         \put(52,1){$\epsilon$}
         \put(-3,22.2){$|\delta V_r|$}
	 \put(-1,42){(B)}
    \end{overpic}
    \begin{overpic}[width=0.83\textwidth]{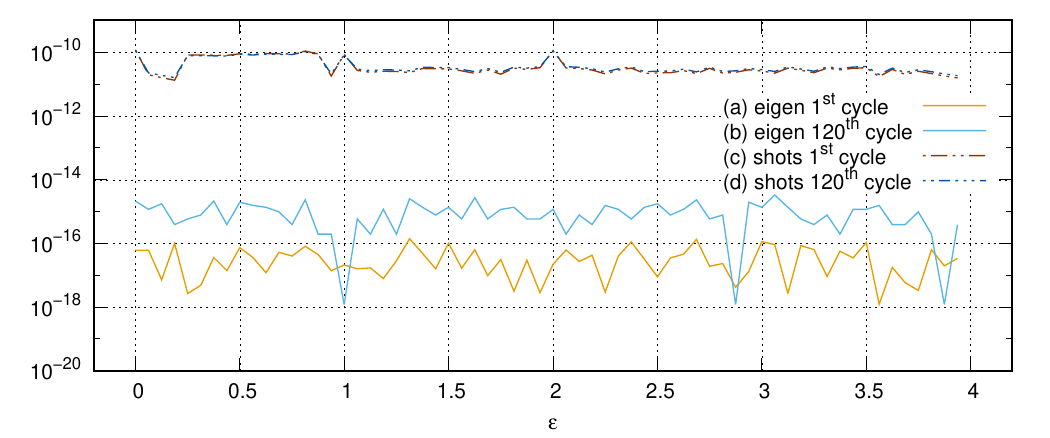}
         \put(52.6,1.6){\makebox(0,0)[c]{\color{white}\rule{1.5em}{2.5ex}}}
         \put(52,1){$\epsilon$}
         \put(-3,22.2){$|\delta V_r|$}
	 \put(-1,42){(C)}
    \end{overpic}
    \begin{overpic}[width=0.83\textwidth]{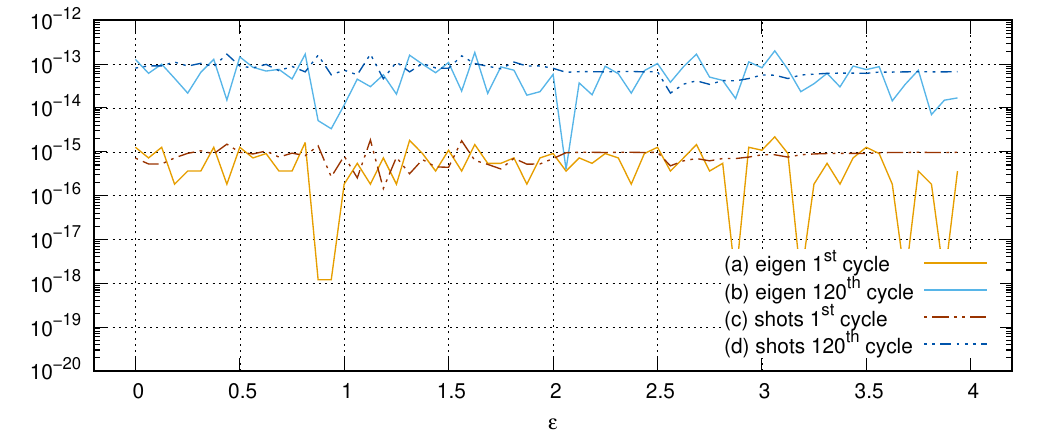}
         \put(52.6,1.6){\makebox(0,0)[c]{\color{white}\rule{1.5em}{2.5ex}}}
         \put(52,1){$\epsilon$}
         \put(-3,22.2){$|\delta V_r|$}
	 \put(-1,42){(D)}
    \end{overpic}
    \caption{%
    The changes in magnitude of the relative differences $|\delta V_r|$ along the evolution presented as representative constant $\phi$ slices of maps presented in Fig.~\ref{fig:bounceMap2} for chosen stages of evolution ($\Lambda=-0.05$, $\omega_o=4000$, $\sigma_{\omega}=200G^{1/2}$): the bounce (A), the expansion phase $(B)$, the recollapse $(C)$ and the contraction $(D)$ respectively. Each subfigure shows $|\delta V_r|$ for the $1^{\rm st}$ evolution cycle (\texttt{eigen} $(a)$ and shots method $(c)$ respectively) and for the $120^{\rm th}$ evolution cycle (\texttt{eigen} $(b)$ and shots method $(d)$ respectively).
    }
    \label{fig:slices1}
\end{figure*}

\begin{figure*}[p]
    \begin{overpic}[width=0.83\textwidth]{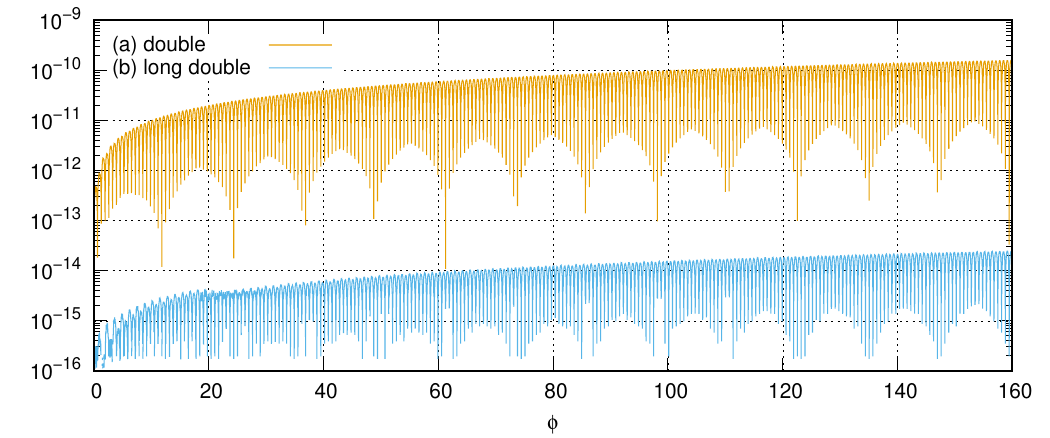}
         \put(-3,22.2){$|\delta V_r|$}
	 \put(-1,42){(A)}
    \end{overpic}
    \begin{overpic}[width=0.83\textwidth]{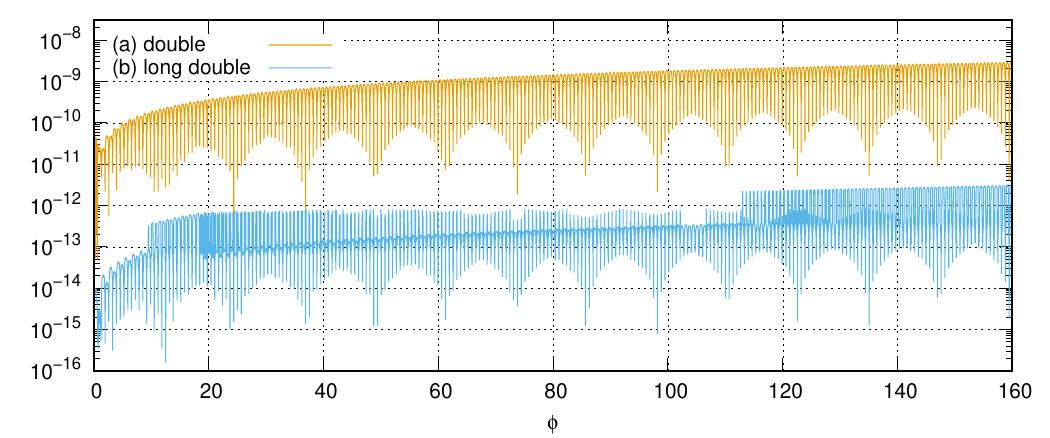}
         \put(-3,22.2){$|\delta \sigma_r|$}
	 \put(-1,42){(B)}
    \end{overpic}
    \caption{A section $\epsilon=2.375$ of the relative differences $|\delta V_r|$ $(A)$ and $|\delta\sigma_r|$ $(B)$ for the example state presented in Fig.~\ref{fig:bounceMap2} built of the basis evaluated via \texttt{eigen} method with \texttt{double} $(a)$ and \texttt{long double} $(b)$ precision respectively. One observes drop in values of both quantities by several levels of magnitude as precision increases.}
    \label{fig:relDiff1}
\end{figure*}

\begin{figure*}[p]
    \begin{overpic}[width=0.83\textwidth]{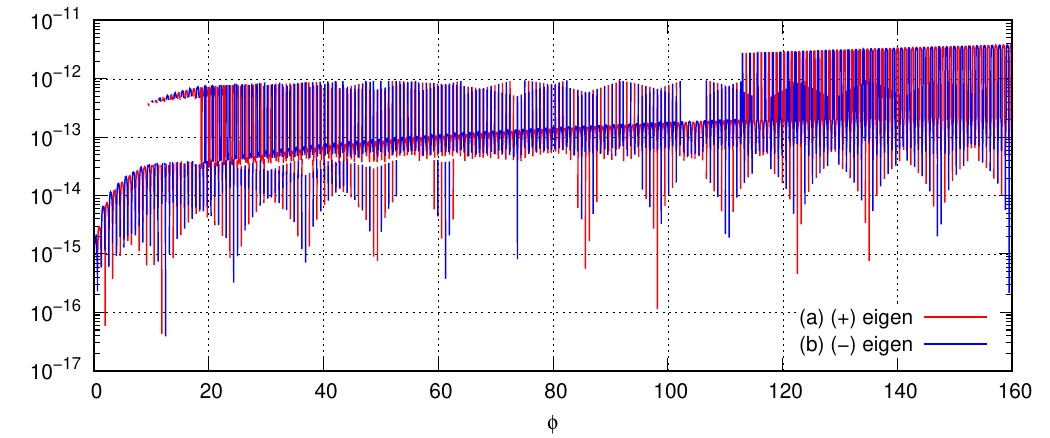}
         \put(-3,22.2){$\overline{\delta\sigma_r}$}
	 \put(-1,42){(A)}
    \end{overpic}
    \begin{overpic}[width=0.83\textwidth]{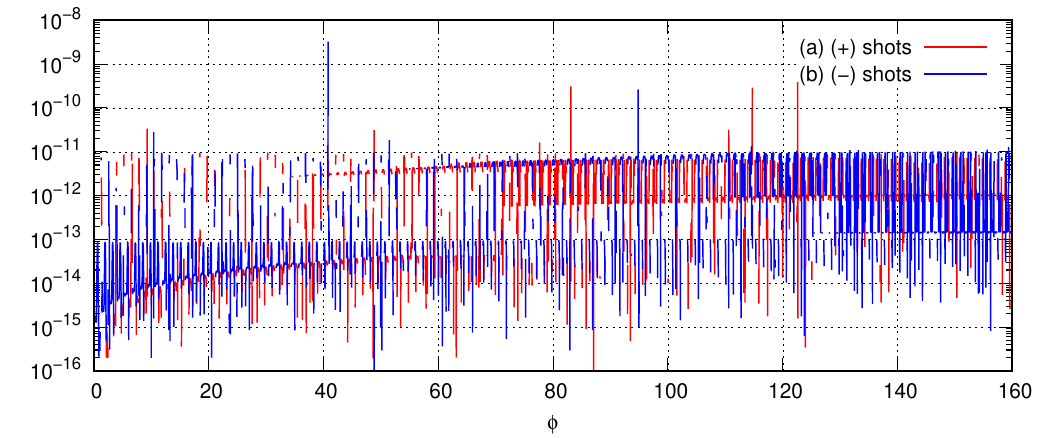}
         \put(-3,22.2){$\overline{\delta\sigma_r}$}
	 \put(-1,42){(B)}
    \end{overpic}
    \begin{overpic}[width=0.83\textwidth]{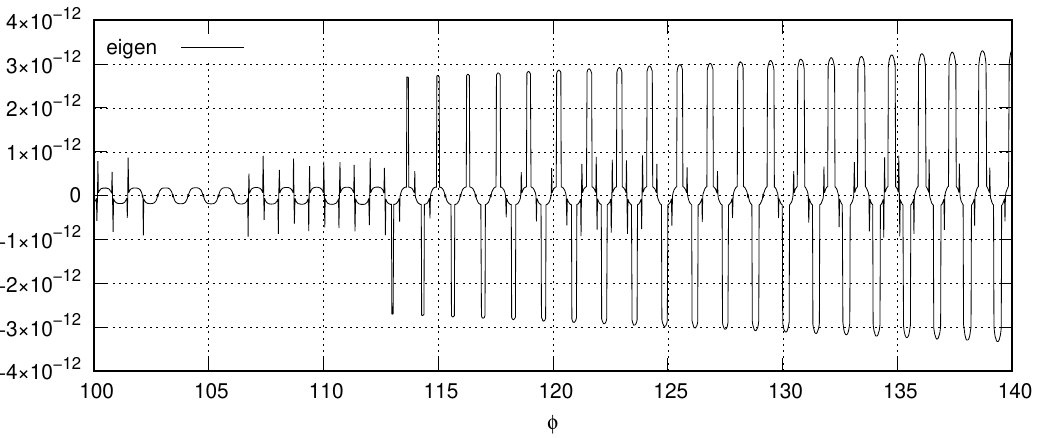}
         \put(-3,22.2){$\overline{\delta\sigma_r}$}
	 \put(-1,42){(C)}
    \end{overpic}
    \begin{overpic}[width=0.83\textwidth]{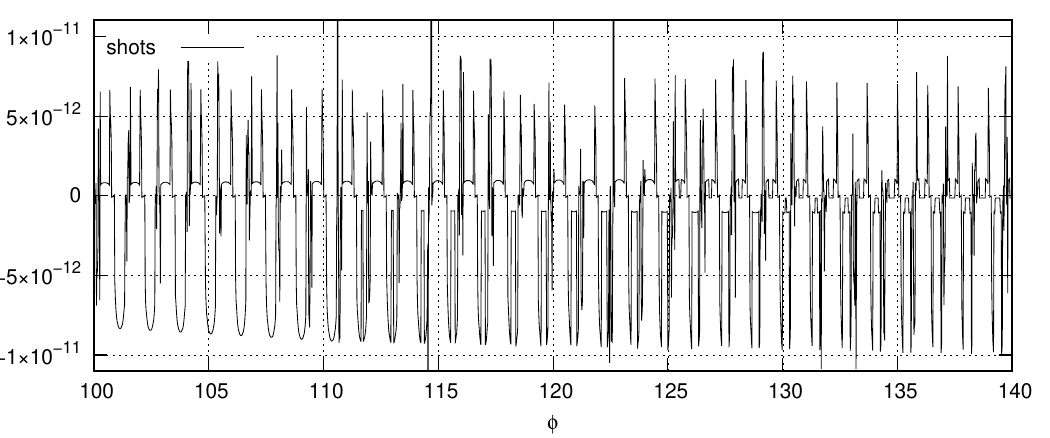}
         \put(-3,22.2){$\overline{\delta\sigma_r}$}
	 \put(-1,42){(D)}
    \end{overpic}
    \caption{Evolution over $\phi$ of the averaged relative difference $\overline{\delta\sigma_r}$ \eqref{eq:avg_sigma} for the example state presented in Fig.~\ref{fig:bounceMap2} formed out of basis generated by \texttt{eigen} $(A,C)$ and shots $(B,D)$ method respectively. In order to illustrate the sign change while keeping the logarithmic scale $|\overline{\delta\sigma_r}|$ is plotted in two separate branches plotted for the domain where it reaches positive $(a)$ and negative $(b)$ values respectively. Subfigures $(C)$ and $(D)$ present the zoom of $(A)$ and $(B)$ respectively (without splitting onto separate branches) in order to illustrate the quasi-oscillatory behavior of probed quantity ($\Lambda=-0.05$, $\omega_o=4000$, $\sigma_{\omega}=200G^{1/2}$).
    }
    \label{fig:relDiffDvSignedHistory1}
\end{figure*}

\begin{figure*}[p]
    \begin{overpic}[width=0.83\textwidth]{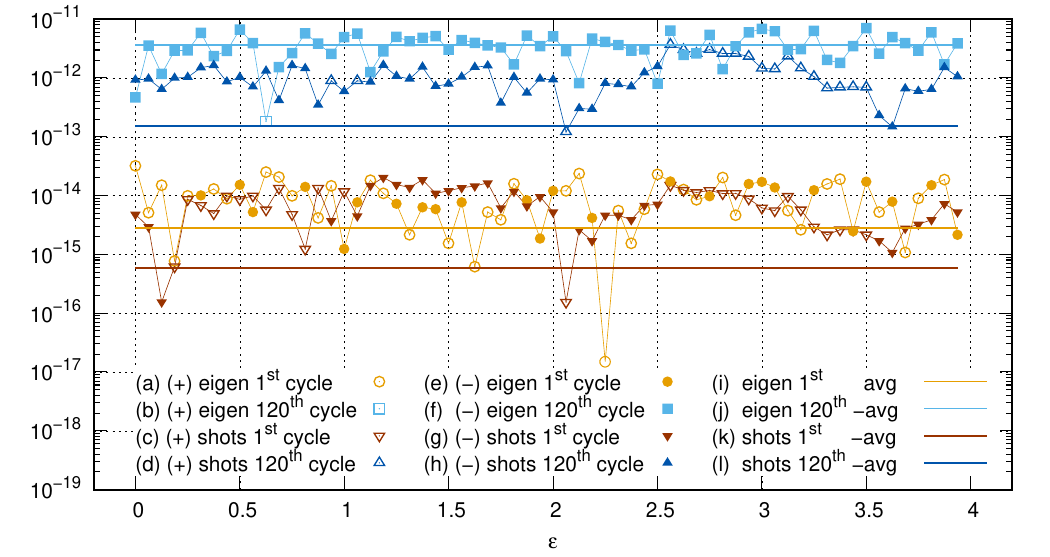}
         \put(52.6,1.6){\makebox(0,0)[c]{\color{white}\rule{1.5em}{2.5ex}}}
         \put(-4.5,28.2){${\delta\sigma_r(\epsilon)}$}
         \put(52,1){$\epsilon$}
    \end{overpic}
    \caption{A comparison of the averaged relative differences $\overline{\delta\sigma_r}$ with their components $\delta\sigma_r(\epsilon)$ (integrand in \eqref{eq:avg_sigma}) for chosen moments of $\phi$ near the beginning and the end of numerically probed evolution of the example state illustrated in Fig.~\ref{fig:bounceMap2}. The profiles correspond to the middle of the expansion phase in the $1^{\rm st}$ $(a,c,e,g)$ and $120^{\rm th}$ $(b,d,f,h)$ cycle of evolution. The state has been composed out of basis generated via two methods: \texttt{eigen} $(a,b,e,f)$ and shots $(c,d,g,h)$ one. We do not observe significant increase between cycles. Also, the results utilizing shots method are much better balanced about zero, being however much less accurate. Again, in order to visualize sign changes the absolute value of the integrand were plotted in two separate branches corresponding to positive (empty point style) and negative values (filled point style). The average $\overline{\delta\sigma_r}$ $(i)$ ($1^{\rm st}$ stage, \texttt{eigen} method) is positive, whereas the remaining ones $(j,k,l)$ are negative, while their absolute values are presented in the plot. 
    }
    \label{fig:relDiffDvSigned}
\end{figure*}

\begin{figure*}[p]
    \begin{overpic}[width=0.83\textwidth]{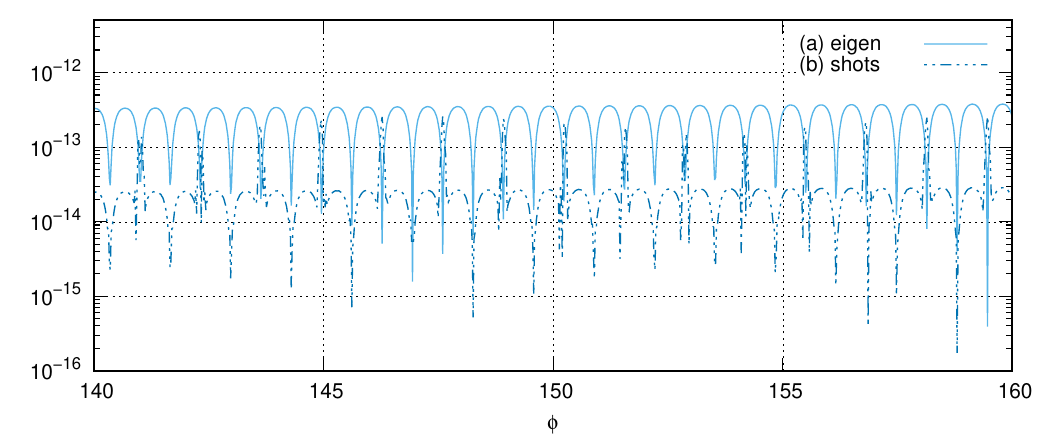}
         \put(-3,22.2){$|\delta V_r|$}
	 \put(-1,42){(A)}
    \end{overpic}
    \begin{overpic}[width=0.83\textwidth]{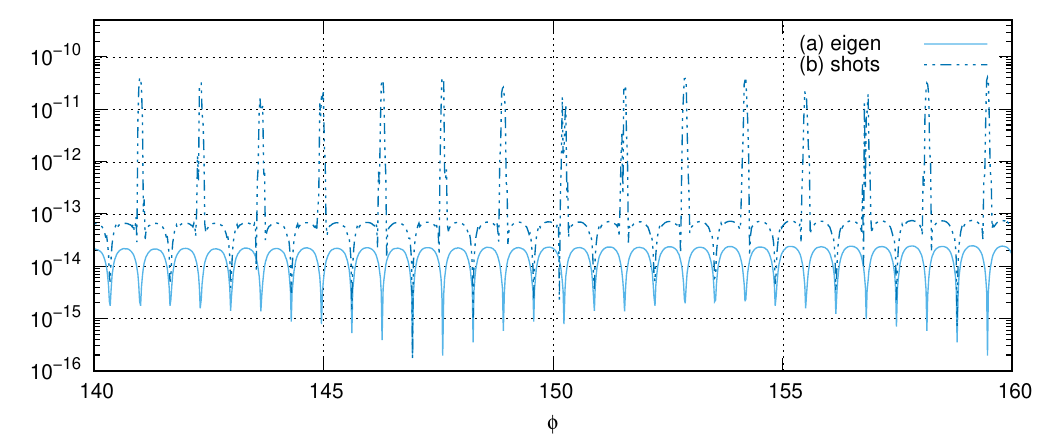}
         \put(-3,22.2){$|\delta V_r|$}
	 \put(-1,42){(B)}
    \end{overpic}
    \begin{overpic}[width=0.83\textwidth]{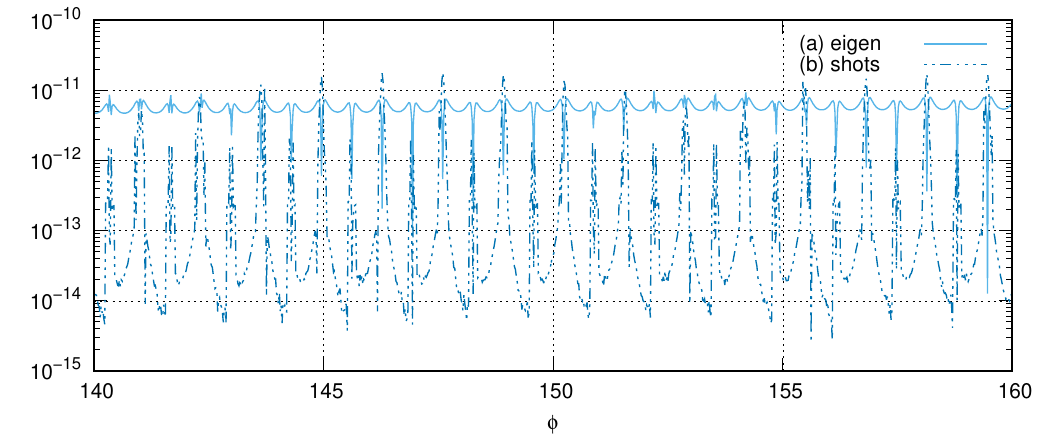}
         \put(-3,22.2){$|\delta \sigma_r|$}
	 \put(-1,42){(C)}
    \end{overpic}
    \begin{overpic}[width=0.83\textwidth]{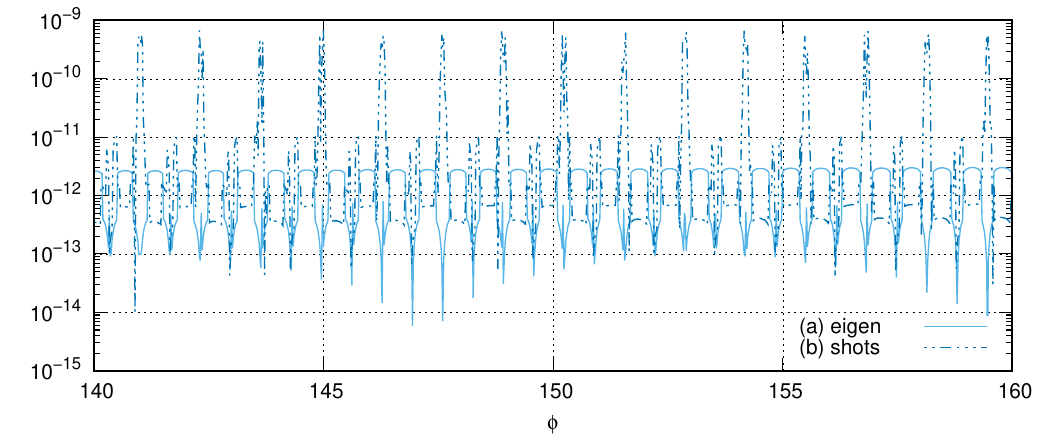}
         \put(-3,22.2){$|\delta \sigma_r|$}
	 \put(-1,42){(D)}
    \end{overpic}
    \caption{%
    An illustration of the evolution of the(absolute value of the) relative differences $\delta V_r$ in (A,B) and $\delta\sigma_r$ in~(C,D) for Gaussian built out of basis generate via \texttt{eigen} $(a)$ and shots $(b)$ method respectively, for superselection sector $\epsilon=2.375$. Subfigures $(A,C)$ and $(B,D)$ show the results for the states peaked about $\omega_o=1000G^{1/2}$ (with $\sigma_{\omega}=50G^{1/2}$) and $\omega_o=4000G^{1/2}$ (with $\sigma_{\omega}=200G^{1/2}$) respectively.
    }
    \label{fig:growingOscillation3}
\end{figure*}

\begin{figure*}[p]
    \begin{overpic}[width=0.83\textwidth]{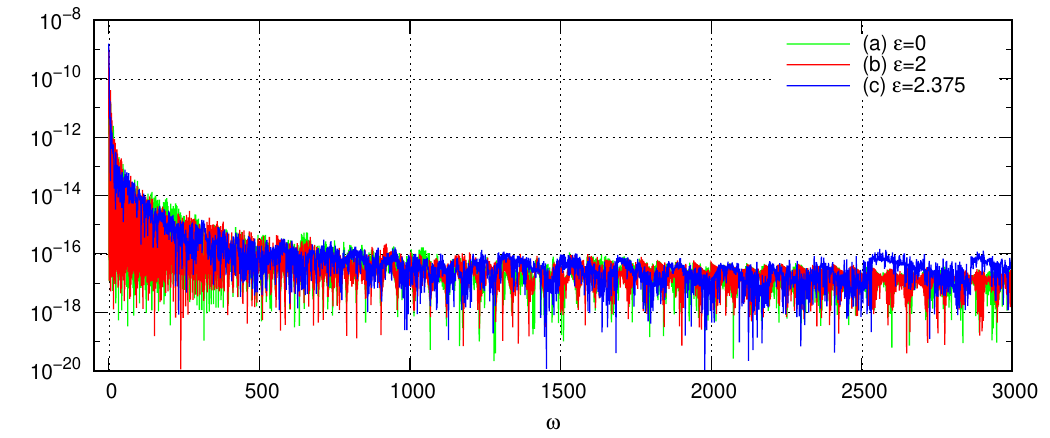}
        \put(53,1.5){\makebox(0,0)[c]{\color{white}\rule{1.5em}{2.5ex}}}
        \put(53,1){$i$}
	\put(-2,22.5){$\delta\omega_i$}
    \end{overpic}
    \caption{Comparison of relative differences $\delta\omega_i=|\omega_{i,\texttt{eigen}}-\omega_{i,\textrm{shots}}|/\omega_{i,\textrm{shots}}$  between the values of spectrum elements of $\sqrt{\hat{\Theta}_{\Lambda}}$ for $\Lambda=-0.05$ found via \texttt{eigen} and shots method is shown for the superselection sector: $\epsilon=0$ (a), $\epsilon=2$ (b) and $\epsilon=2.375$ (c) respectively.}
    \label{fig:eigenvaluesComparison1}
\end{figure*}

\begin{figure*}[p]
    \begin{overpic}[width=0.83\textwidth]{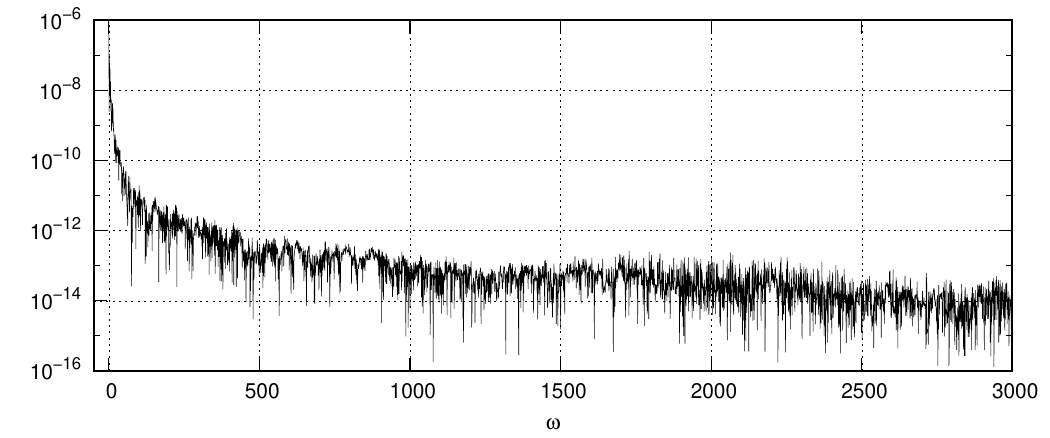}
        \put(53,1.5){\makebox(0,0)[c]{\color{white}\rule{1.5em}{2.5ex}}}
        \put(53,1){$i$}
	\put(-3,22.5){$\delta\omega_{i,\texttt{ld}}$}
    \end{overpic}
    \caption{Comparison of relative error of eigenvalues in \texttt{double} and \texttt{long double} precisions calculated by \texttt{eigen} library: $\delta\omega_{i,\texttt{ld}}=|\omega_{i,\texttt{double}}-\omega_{i,\texttt{long double}}|/\omega_{i,\texttt{long double}}$ (logarithmic scale).
    }
    \label{fig:eigenvaluesComparison2}
\end{figure*}

In order to analyze the differences in the trajectories/variances the integral ones were used as a reference point and the single sector ones have been compared against them. Specifically the behavior of 
the relative differences 
\begin{equation}
    \delta V_r(\epsilon,\phi) := (\langle V \rangle_\epsilon - \langle V \rangle)/\langle V \rangle \ , 
    \qquad
    \delta\sigma_r(\epsilon,\phi) := (\Delta V_\epsilon - \Delta V)/\Delta V \ 
    \label{eq:RelDiffDef}
\end{equation}
has been analyzed. Such comparison allows to easily detect the differences between distinct superselection sectors. Furthermore, any distinct feature of the integral states (like stronger decoherence) would show up as a trend (relative variation difference increasing along the evolution in $\phi$) in the data listed. The analysis of $(\delta V_r,\delta\sigma_r)$ has been performed for the population of the states specified in the previous section with use of both methods of calculating the Hilbert space basis and subsequently compared. Furthermore, the data analysis has been supplemented by evaluating $(\delta V_r,\delta\sigma_r)$ using the basis calculated with lower precision: using \texttt{eigen} library with \texttt{double} precision, whereas all other calculations used \texttt{long double} precision. The results of this analysis are presented on Figs.~\ref{fig:bounceMap2} -- \ref{fig:eigenvaluesComparison2} and are:
\begin{enumerate}
  \item The relative differences $(\delta V_r,\delta\sigma_r)$ remained smaller than $10^{-9}$ throughout the evolution (performed within the range $\phi\in [0,160]G^{-1/2}$) independently of the method used, as it can be seen in Fig.~\ref{fig:bounceMap2}. There, the left part of Fig.~\ref{fig:bounceMap2} shows the expectation value $\langle V\rangle$ over several bounce-recollapse cycles, the middle colormap shows the relative difference $\delta V_r(\epsilon,\phi)$ and the right colormap shows $\delta\sigma_r(\epsilon,\phi)$. The top row corresponds to calculations using the \texttt{eigen} library and the bottom row corresponds to calculations using the shots method. 
  \item \label{it:decoherence} As the state evolves ($\phi$ increases) one observes a slight increase of $|\delta V_r|$ and $|\delta\sigma_r|$, however the comparison of the results calculated with use of basis generated with a different precision shows, that this increase should be attributed to the numerical error. This is shown in Fig.~\ref{fig:relDiff1}, where the evaluation of the relative differences $\delta V_r$ and $\delta\sigma_r$ with use of basis evaluated with different precision (\texttt{eigen} library: \texttt{double} vs. \texttt{long double}) is presented for an example of a Gaussian state. There, upon increasing the precision (here by about 5 levels of magnitude) the evaluated quantities drop, in presented case by 3 to 5 levels of magnitude.
  \item When the basis is generated by \texttt{eigen} library (top row in Fig.~\ref{fig:bounceMap2}) the relative differences are in fact smaller than $10^{-11}$, while the old shots method (bottom row in Fig.~\ref{fig:bounceMap2}) has visible sharp peaks (the yellow-green horizontal lines on both colormaps) corresponding to points of recollapse, which amounts to increase of the differences to $10^{-9}$.
  A vertical cross-section of the middle colormap for a single sector $\epsilon=2.375$ shows this phenomenon in Fig.~\ref{fig:growingOscillation3} -- the sharp peaks for the shots method are clearly distinguished. What is remarkable, once these spikes are excluded, the maximum error for the shots method drops below $10^{-13}$~(Fig.~\ref{fig:growingOscillation3}B).
  
  Furthermore, as one can observe in Fig.~\ref{fig:growingOscillation3}, the shots method becomes less accurate as the value of $\omega_o$ increases, while \texttt{eigen} library keeps the error approximately similar. In particular, for $\omega_o\approx 10^3G^{1/2}$ the shots method is actually more accurate. The former is a consequence of the iterative method of evaluating the generalized  eigenfunctions.
  
  The peaks discussed above are generated due to an oversimplified method of extrapolating the decaying tails of basis functions, where the simple extrapolation via exponential decay past the recollapsing point $v_r$ has been implemented in the code. A constructive interference of the errors due to extrapolation error shows up exactly as a peak at the recollapse. A more precise extrapolation using eigenstates of the evolution operator of the same model in geometrodynamics (Wheeler-DeWitt) framework would reduce the error by several levels of magnitude. However as they take the form of a Bessel function of the third kind of imaginary order (as discussed earlier) this would significantly increase the numerical difficulty and was not implemented due to limited use. 
  \item Across all $\epsilon$ sectors the relative difference of volume stays almost the same, within the range of numerical error, as shown on Figure~\ref{fig:slices1}, which shows a horizontal cross section of these colormaps done at bounce (Fig.~\ref{fig:slices1}A), in the middle of expansion (Fig.~\ref{fig:slices1}B), at recollapse (Fig.~\ref{fig:slices1}C) and in the middle of contraction (Fig.~\ref{fig:slices1}D). Across all sectors the value stays the same within two orders of magnitude, well below $10^{-12}$ with the exception of the data generated via shots method at recollapse, where the error spikes to $10^{-10}$, as discussed earlier.
  \item The integral states do not show any detectable increase in variance of volume throughout the evolution in comparison to single sector ones. Indeed, the analysis of the averaged relative difference
  \begin{equation}\label{eq:avg_sigma}
    \overline{\delta\sigma_r}(\phi) := \frac{1}{4}\int_0^4 \delta\sigma_r(\epsilon,\phi)\rd\epsilon     
  \end{equation}
  shows no particular trend in its behavior. Have the integral state spread faster than its single sector components, $\overline{\delta\sigma_r}$ would consistently take increasingly negative values as the state evolves. However no such behavior has been identified in the analysis of the population of states probed. An example of this analysis is shown on Figs.~\ref{fig:relDiffDvSignedHistory1} and~\ref{fig:relDiffDvSigned}. One observes, that $\overline{\delta\sigma_r}$ changes sign in a quasi-periodic manner, consistently oscillating around zero (see Fig.~\ref{fig:relDiffDvSignedHistory1}). Its behavior in relation to its integrand $\delta\sigma_r$ is a bit different depending on the method of generating the basis (see Fig.~\ref{fig:relDiffDvSigned}): $(i)$ for the basis generated via \texttt{eigen} both quantities are of the same order, whereas $(ii)$ for the old shots method the average takes significantly lower values than the integrand. While in principle $(i)$ could lead to a trend, the quasi-oscillatory nature of that apparent trend shows, that within precision of the methods used no increased decoherence of the integral states is visible.
  \item The results computed with use of bases generated via different methods are consistent. Indeed, the comparison of the eigenvalues of $\hat{\Theta}_{\Lambda}$ evaluated by \texttt{eigen} library and shots method shows an agreement to within the $10^{-14}$ relative difference. This can be seen in Fig.~\ref{fig:eigenvaluesComparison1}, which shows the relative difference between the first 3000 eigenvalues for three sectors $\epsilon\in\{0, 2, 2.375\}$. 
  
  Furthermore, in computing the spectrum of $\hat{\Theta}_{\Lambda}$ to within  $10^{-12}$ accuracy it is enough to employ the \texttt{eigen} library with a standard \texttt{double} precision. Indeed, fig.~\ref{fig:eigenvaluesComparison2} shows the comparison of \texttt{long double} (18 decimal places) vs. \texttt{double} (15 decimal places) for \texttt{eigen} library, confirming this observation. Note however, that this accuracy is not sufficient for determining the basis, as it generates a visible spurious decoherence when that basis is used in probing the semiclassical state dynamics (see the discussion in point.~\ref{it:decoherence}).
\end{enumerate}

From the above observations we can draw several general conclusions regarding this particular model:
\begin{itemize}
  \item {\bf Robustness:} The properties of the ``energy'' eigenstates and semiclassical states reported originally in \cite{Bentivegna2008} are confirmed. In particular there are relevant differences between the superselection sectors. Furthermore, the old method of shots used in \cite{Bentivegna2008} is positively tested by the independent one employing \texttt{eigen} library.
  \item {\bf Properties of integral states:} Within the precision of the methods used the integral states preserve the properties of the single sector ones. In particular there is no noticeable increase of decoherence due to differences between sectors.
  \item {\bf Methodology:} A simple application of a widely used standard numerical library (\texttt{eigen}) yields no less accurate (and in many cases even more accurate) results than using a relatively involved code dedicated to this class of problems. However, due to its limitations (finite operator matrices) this comes at a cost of a significant (levels of magnitude) increase of computational resources required.
\end{itemize}

\begin{figure*}[p]%%%%%%%%%%%%%%%%%%% ↓ , grid, tics=2
    \begin{overpic}[width=0.83\textwidth]{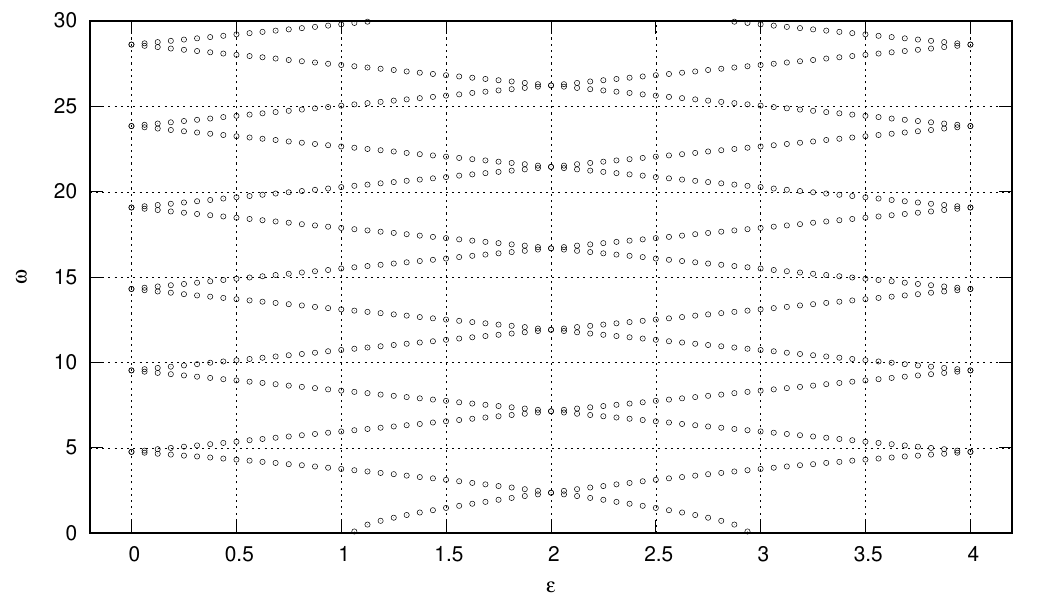}
        \put(52,1){\makebox(0,0)[c]{\color{white}\rule{1.5em}{2.5ex}}}
        \put(52,1){${\epsilon}$}%%%%%%%% ↑ yellow
        \put(1.5,30.5){\makebox(0,0)[c]{\color{white}\rule{1.5em}{2.5ex}}}
        \put(1.5,30.5){${\omega}$}
    \end{overpic}
    \caption{An example of a dependence of the lowest eigenvalues of $\sqrt{\hat{\Theta}_{\Lambda}}$ operator on the superselection sector, here shown for $\Lambda=-0.05\lPl^{-2}$. One observes a continuous dependence on $\epsilon$. Consequently, any positive real value is an eigenfunction of $\sqrt{\hat{\Theta}_{\Lambda}}$ for some $\epsilon$.}
    \label{fig:Eigenvalues}
\end{figure*}

\begin{figure*}[p]
    \begin{overpic}[height=0.9\textwidth,angle=270]{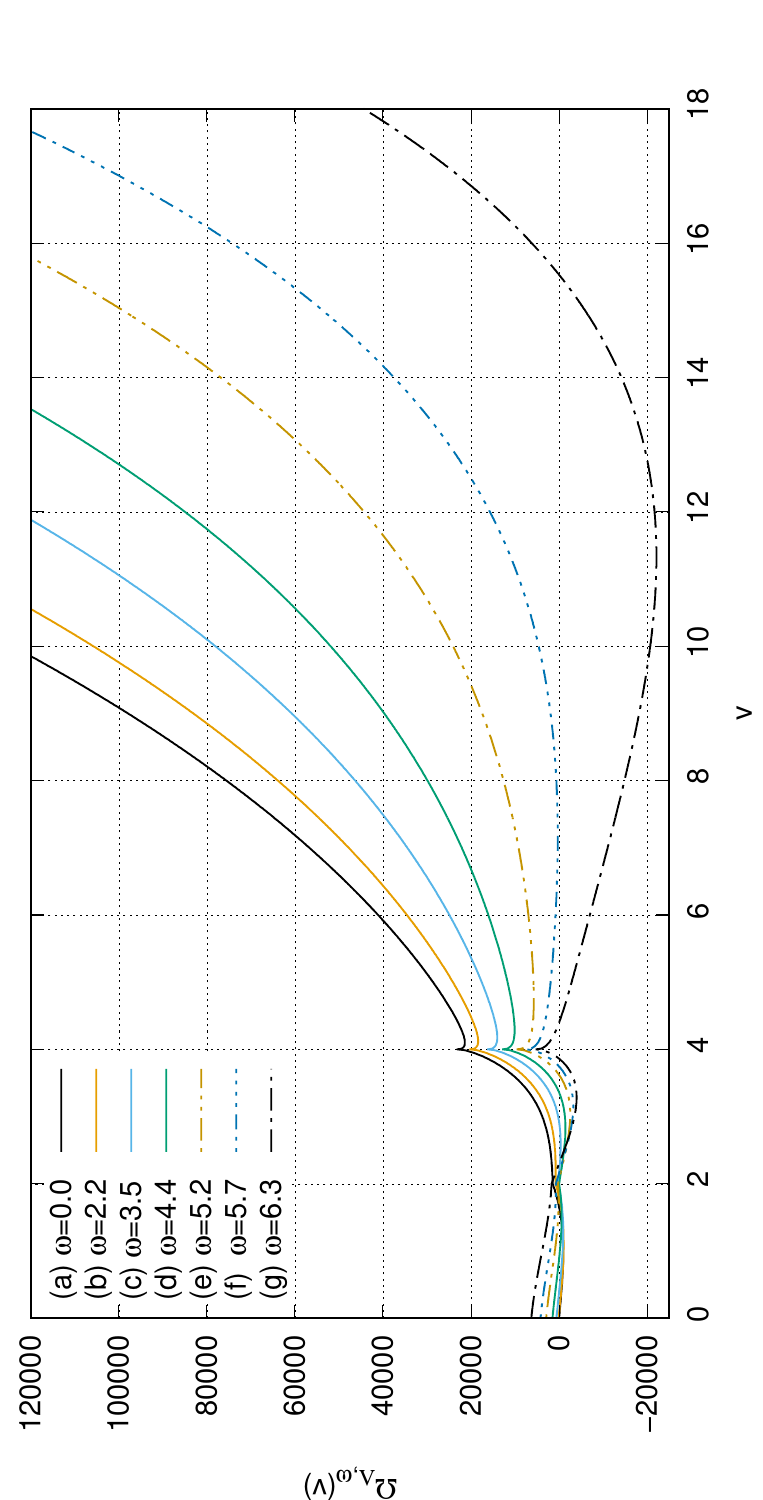}
        \put(52,1.5){\makebox(0,0)[c]{\color{white}\rule{1.5em}{2.5ex}}}
        \put(51.5,1.0){$v$}
        \put(1.3,27){\makebox(0,0)[c]{\color{white}\rule{1.5em}{9.5ex}}}
	\put(-3.0,27.5){$\Omega_{\Lambda,\omega}(v)$}
    \end{overpic}
	\caption{Function $\Omega_{\Lambda,\omega}(v)$ \eqref{eq:Omega-def} for $\Lambda=-0.05$ is plotted for several values of $\omega$:
	(a)~$0.0$;
	(b)~$2.2$;
	(c)~$3.5$;
	(d)~$4.4$;
	(e)~$5.2$;
	(f)~$5.7$ and
	(g)~$6.3$.
	Units: $G^{1/2}$.}
	\label{fig:OmegaFunc}
\end{figure*}

\begin{figure*}[p]
    \begin{overpic}[height=0.49\textwidth,angle=270]{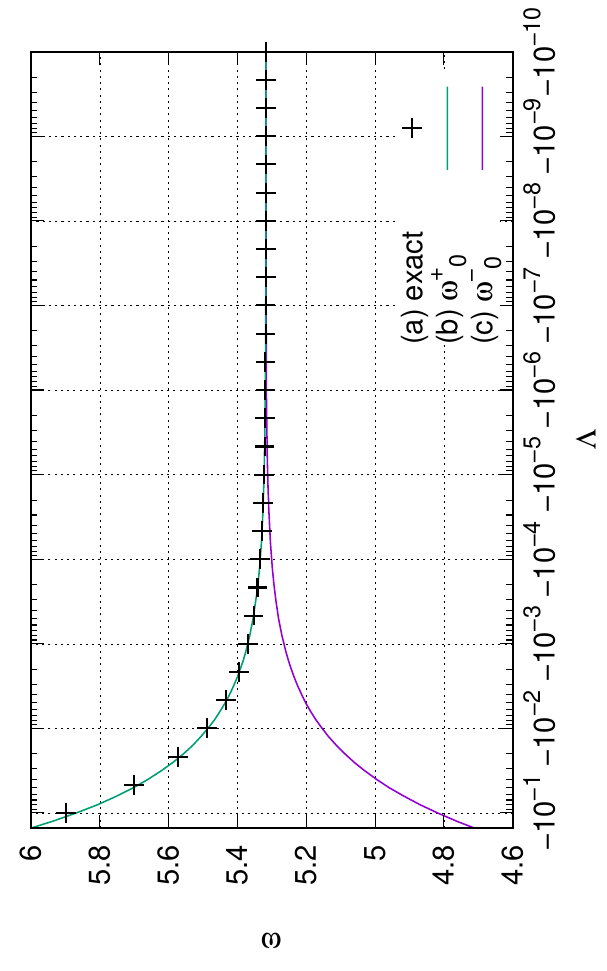}
        \put(52.6,1.6){\makebox(0,0)[c]{\color{white}\rule{1.5em}{2.5ex}}}
        \put(52,1){$\Lambda$}
        \put(2,34){\makebox(0,0)[c]{\color{white}\rule{1.5em}{2.5ex}}}
        \put(1,33){$\omega$}
        \put(-1,62){(A)}
    \end{overpic}
    \begin{overpic}[height=0.49\textwidth,angle=270]{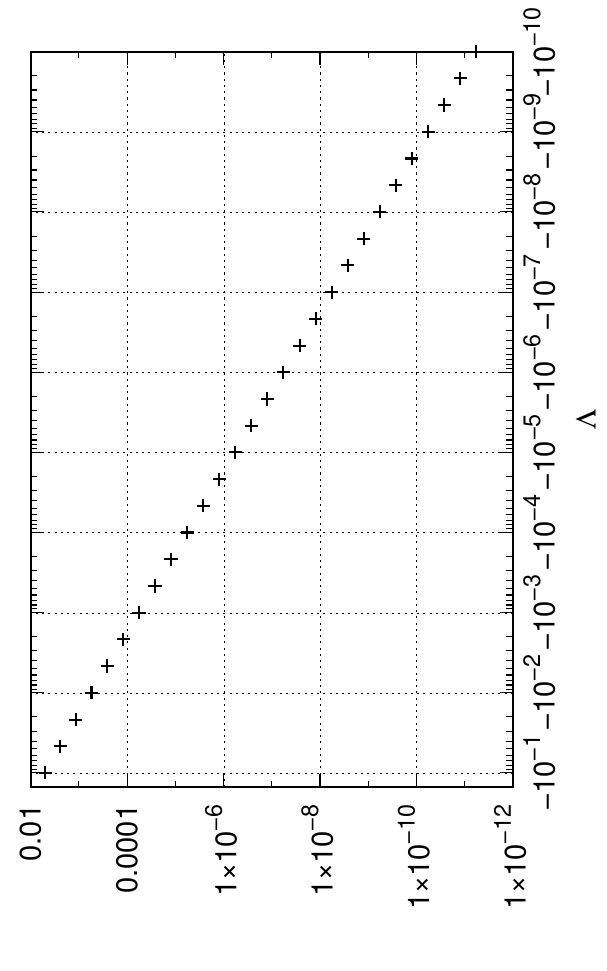}
        \put(55.6,1.6){\makebox(0,0)[c]{\color{white}\rule{1.5em}{2.5ex}}}
        \put(55,1){$\Lambda$}
        \put(-1,62){(B)}
	\put(-1,34.0){$\delta\omega^\pm$}
    \end{overpic}
    \caption{Approximate solutions $\omega^{\pm}_0$ given via \eqref{eq:omega-transition-approx} are compared against the actual numerical solutions to Eq.~\ref{eq:systemOmega} (A). One sees that only $\omega^{+}_0$ approximates the actual solution relatively well. The accuracy (relative difference) {${\delta\omega^\pm = |\textrm{exact}-\omega^+_0|/\textrm{exact}}$} of this approximation is shown in (B).
    }
	\label{fig:omegaVsLambda}
\end{figure*}

\section{Conclusions}
\label{sec:conclusions}

In presented work we compared properties of semiclassical states representing an isotropic FLRW universe quantized within LQC framework and constructed as states within integral versus single superselection sector physical Hilbert spaces. For that comparison we choose the model of an isotropic universe admitting negative cosmological constant and massless scalar field -- a relatiely simple model featuring a quasi-periodic evolution through a chain of bounces and recollapses, which is very convenient for studying long term evolution effects. The studies focused on search for minuscule differences between integral and single sector states as well as between the superselection sectors. To increase the robustness of the studies the physical Hilbert space basis used for constructing the states was identified via two distinct numerical methods: the shots method originally used in the previous studies of the system and a new one directly applying standard numerical tools, in this case the \texttt{eigen} library. The main questions posed here were: $(i)$ do integral states have worse semiclassical properties than single sector ones, and $(ii)$ is the dedicated code for finding the Hilbert space basis used originally in \cite{Bentivegna2008} robust and can it be replaced by standard tools?

On one hand, $(i)$ has been answered in the negative: while high precision numerical analysis shows differences between listed classes of states, they do not amount to a detectable loss of semiclassical properties of the integral states with respect to single sector ones. Consequently the integral states can safely replace the projections onto superselection sectors and the conclusions regarding those states' properties remain in power. This result opens a very promissing avenue of analyzing the dynamics of more complicated systems (starting from i.e. Bianchi I models), where due to the nature of superselection sectors themselves (singular measure spaces) use of single sector states is not technically viable. One has to remember however, that this is a result established for one particular model and does not automatically apply to other ones. Specifically, for the model considered the spectrum of energy eigenstates (forming the preferred basis) for each superselection sector consists of countable sets of isolated points, whereas the integral over superselection sectors admits whole $\re^+$ as a set of possible energy eigenvalues. It is a priori not known whether the integrals of the sectors already admitting continuous spectra will be equally free from problems. This remains a question for further studies.

On the other hand, $(ii)$  has been answered in the affirmative: the new method of calculating the basis confirmed the correctness of the previous one but also in most cases exceeded it in precision. Consequently, we have in our hands a viable numerical tool that is much easier to apply. This comes however at a cost of computational resources, as the new method requires at least one order of magnitude more of them.

\begin{acknowledgments}
This work was supported in part by the Polish National Center for Science (Narodowe Centrum Nauki – NCN) grant OPUS 2020/37/B/ST2/03604.
The calculations were performed, in part, using the resources of Division of Theoretical Physics and Quantum Information, Faculty of Applied Physics and Mathematics at Gdańsk University of Technology.
\end{acknowledgments}

\appendix

\section{Stability of the eigenvalue problem}
\label{app:stability}
In order to understand the qualitative properties of the eigenfunctions of the evolution operator $\Theta_{\Lambda}$ defined in \eqref{eq:theta-Lambda} it is good to recast the initial value problem for the generalized eigenfunctions in $1$st order form. To do so, lets consider a general eigenfunction $\psi_{\omega}$ corresponding to the eigenvalue $\omega^2$. Such eigenfunction is determined as a solution to \eqref{eq:2ndOrder} (with coefficients $f_{\pm,o}$ given by \eqref{eq:2ndOrder-coeff}). 
By introducing the representation of $\psi_{\omega}$ as a vector
\begin{equation}
  \vec{\psi}_{\omega}(v) = \left( \begin{array}{c} \psi(v) \\ \psi(v-4) \end{array} \right) 
\end{equation}
one can rewrite the (implied by \eqref{eq:2ndOrder}) relation between consecutive values of $\psi_{\omega}$ at some lattice $\lat_{\epsilon}$ as an iterative step
\begin{equation}\label{eq:eigen-iter-pos}
  \vec{\psi}_{\omega}(v+4)  
  = \left( \begin{array}{cc} A(v) & B(v) \\ 1 & 0 \end{array} \right) \vec{\psi}_{\omega}(v) 
  =: M^+(v) \vec{\psi}_{\omega}(v) \ , 
\end{equation}
where
\begin{equation}
  A(v) = \frac{f_o(v)-\pi G\gamma^2\Delta\Lambda v^2-\omega^2}{f_+(v)} \ ,
  \quad
  B(v) = -\frac{f_-(v)}{f_+(v)} \ .
\end{equation}
Analogously one can define a vector representation 
\begin{equation}
  \vec{\phi}(v) = \left(\begin{array}{c}\psi(v) \\ \psi(v-4)\end{array}\right) \ ,
\end{equation}
and subsequently write down the iterative relation between consecutive values of $\psi$ as
\begin{equation}\label{eq:eigen-iter-neg}
  \vec{\phi}_{\omega}(v-4)  
  = \left( \begin{array}{cc} \tilde{A}(v) & \tilde{B}(v) \\ 1 & 0 \end{array} \right) \vec{\psi}_{\omega}(v) 
  =: M^-(v) \vec{\psi}_{\omega}(v)  \ , 
\end{equation}
where
\begin{equation}
  \tilde{A}(v) = \frac{f_o(v)-\pi G\gamma^2\Delta\Lambda v^2-\omega^2}{f_-(v)} \ ,
  \quad
  \tilde{B}(v) = -\frac{f_+(v)}{f_-(v)} \ .
\end{equation}
The behavior of the solution $\vec{\psi}_{\omega}(v)$ is governed by the spectral properties of the transfer matrix $M^+$. It is quite easy to find the eigenvalues of the latter as functions of $\Lambda$, $\omega$ and $v$. They are
\begin{equation}\label{eq:M-eigenvals}
\delta_\pm=\frac{\pi  G v^2 \left(3-2 \gamma ^2 \Delta  \Lambda \right)-2 \omega ^2\pm\sqrt{\left(\pi  G v^2 \left(2 \gamma ^2 \Delta  \Lambda -3\right)+2 \omega ^2\right)^2-9 \pi ^2 G^2 \left| \sqrt{v-4} v \sqrt{v+4} \left(v^2-4\right)\right| }}{3 \pi  G | v+2|  \sqrt{| v (v+4)| }}
\end{equation}
From this formula one immediately sees, that they are either purely real or form a pair of mutually adjoint complex quantities. In the former case the eigenvalue experiences quasi-exponential behavior, whereas in the latter the behavior is oscillatory. Furthermore, for large $v$ the eigenvalues approach constant asymptotes
\begin{equation}
  \lim_{v\to\pm\infty} \delta^{\pm} = \frac{1}{3}\left[ 3-2\gamma^2\Delta\Lambda \pm \sqrt{(3-2\gamma^2\Delta\Lambda)^2-9} \right] =: \bar{\delta}^{\pm}\ .
\end{equation}
In particular $\bar{\delta}^+ > 1$ whereas $0<\bar{\delta}^-<1$, which indicates both growing and decaying (quasi)exponential mode for sufficiently large $v$.

Let us now identify the zones (in $v$) of particular behavior for a fixed value of $\Lambda$ and $\omega$. They are determined by the sign of the square root argument in \eqref{eq:M-eigenvals}
(an analog of the discriminant)
\begin{equation}\label{eq:Omega-def}
  \Omega_{\Lambda,\omega}(v) := \left(\pi  G v^2 \left(2 \gamma ^2 \Delta  \Lambda -3\right)+2 \omega ^2\right)^2-9 \pi ^2 G^2 \left| \sqrt{v-4} v \sqrt{v+4} \left(v^2-4\right)\right| 
\end{equation}
While the transition points $\Omega_{\Lambda,\omega}(v) = 0$ can be found analytically, their form is excessively complicated, thus in order to analyze its behavior we resort to numerics and analytical approximations. By direct inspections (for wide population of the values of $\Lambda$ and $\omega$ (see Fig.~\ref{fig:OmegaFunc}) we observe that:
\begin{itemize}
  \item In the near-singularity area $|v| < 4$ the behavior of $\Omega_{\Lambda,\omega}(v)$ is relatively complicated -- it changes sign for a part of that domain. However, for each superselection sector only at most two points of the eigenfunction support lie in this area. Consequently it has no significant impact on the eigenfunction behavior.
  \item At $|v| = 4$ we observe a singularity, which reflects the specifics of the initial value problem for the superselection sector $\epsilon=0$ (decoupling of $v=0$ point) and also does not alter the properties of the eigenfunction.
  \item In the domain $|v|>4$ we observe quite tempered changes of $\Omega$: for sufficiently small $\omega$ it grows monotonously, however as $\omega$ grows it starts featuring a minimum, which (for sufficiently large $\omega$) reaches below zero and consequently forms a compact interval where $\Omega_{\Lambda,\omega}$ is negative.
\end{itemize}
In order to characterize in more precise manner the boundaries separating exponential and oscillatory behavior we implement a large $v$ approximation by expanding non-polynomial terms of \eqref{eq:Omega-def}. Since for $v\in\mathbb{R} \setminus (-4,4)$
\begin{equation}
  v\sqrt{v-4}\sqrt{v+4} 
  = v^2 \sqrt{1-\frac{16}{v^2}} 
  = v^2\left(1- \frac{8}{v^2}-\frac{32}{v^4} \right) + O(v^{-4}) \ ,
\end{equation}
the term under consideration can be written as
\begin{equation}
  \left| \sqrt{v-4} v \sqrt{v+4} \left(v^2-4\right)\right| = v^2|v^2 - 12| + O(v^{-2}) \ ,
\end{equation}
which in turn allows one to approximate $\Omega$ by a 2nd order polynomial in $v^2$
\begin{subequations}\begin{align}
  \Omega_{\Lambda,\omega}(v) &= \tilde{\Omega}_{\Lambda,\omega}(v) + O(v^{-2}) \ , 
  \\
  \tilde{\Omega}_{\Lambda,\omega}(v) &:=
  v^4 \left(\pi ^2 G^2 \left(2 \gamma ^2 \Delta  \Lambda -3\right)^2-9 \pi ^2 G^2\right)
  + v^2 \left(108 \pi ^2 G^2+4 \pi  G \omega ^2 \left(2 \gamma ^2 \Delta  \Lambda -3\right)\right)+4 \omega ^4
\end{align}\end{subequations}

For $\Lambda<0$ the leading order coefficient is strictly positive, thus the sign of $\tilde{\Omega}_{\Lambda,\omega}(v)$ depends on the value of the discriminant
\begin{equation}
  \tilde{\Delta}(\Lambda,\omega) := (12\pi G)^2 \left(\omega^4 - 6\pi G (3-2\gamma^2\Delta\Lambda)\omega^2 +(9\pi G)^2 \right) \ ,
\end{equation}
which in turn is a 2nd order polynomial in $\omega^2$. If $\tilde{\Delta}(\Lambda,\omega)>0$ the eigenvalue will admit the oscillatory domain, whereas if $\tilde{\Delta}(\Lambda,\omega)<0$ it will be purely exponential-like. It is quite easy to find its roots, which then will distinguish the transition between two classes of eigenfunctions listed above. They are:
\begin{equation}\label{eq:omega-transition-approx}
  \left(\omega_o^{\pm}(\Lambda) \right)^2  = 3 \pi  \left(\pm 2 \sqrt{\gamma ^4 \Delta ^2 G^2 \Lambda ^2-3 \gamma ^2 \Delta  G^2 \Lambda }-2 \gamma ^2 \Delta  G \Lambda +3 G\right) \ .
\end{equation}
For $\Lambda<0$ we get two valid roots. Note however, that the transition values can also be obtained by numerically solving the system of equations 
\begin{equation}\label{eq:systemOmega}
  \Omega_{\Lambda,\omega}(v) = 0 = \partial_v \Omega_{\Lambda,\omega}(v)
\end{equation}
for $(\omega,v)$. In actual simulations this was done by a builtin Mathematica function \texttt{FindRoot} with the initial values of $\omega$ provided by the solutions \eqref{eq:omega-transition-approx}. The results can be seen in Fig.~\ref{fig:omegaVsLambda}. The non-approximate $\Omega$ features only one transition $\omega_o(\Lambda)$, which is well approximated by the larger solution $\omega_o^+(\Lambda)$. The lower solution $\omega_o^-(\Lambda)$ is then just an artifact of the approximation.

Consequently, for a given value of $\Lambda<0$ the generalized eigenfunctions are purely exponential-like for $0<\omega<\omega_o(\Lambda)\approx\omega_o^+(\Lambda)$ and admit a compact oscillatory domain for $\omega>\omega_o(\Lambda)$. The boundaries of this domain are (in good approximation) given by the roots of $\tilde{\Omega}_{\Lambda,\omega}(v)$
\begin{equation}\label{eq:v-roots}
    (v^{\pm})^2 = \frac{2\omega^4}{\pi} \left( G\omega^2(3-2\gamma^2\Delta\Lambda) - 27\pi G^2 \mp 3G \sqrt{(9\pi G)^2 - 6\pi G(3-2\gamma^2\Delta\Lambda)\omega^2 + \omega^4} \right)^{-1}
\end{equation}

In order to understand the physical meaning of $v^{\pm}$ let us consider a semiclassical state (for example the one defined by (\ref{eq:timeEvol},\ref{eq:projection})) peaked very sharply about a large $\omega$. Then the value $v^-$ corresponds (modulo higher order quantum corrections, i.e. variances etc.) to the expectation value $v(\phi)$ at the bounce, whereas $v^+$ corresponds to the expectation value $v(\phi)$ at the recollapse. For that reason it is convenient to denote the roots as $v_b$ and $v_r$ respectively.

Let us now consider a large $\omega$ limit of the formula \eqref{eq:v-roots}. By taking the leading order of $\omega$ we get
\begin{equation}\label{eq:vbr-app}
  v_b := v^- \approx \frac{\sqrt{2}\omega}{\sqrt{\pi G (6-2\gamma^2\Delta\Lambda)}} \ ,
  \qquad 
  v_r := v^+ \approx \frac{\omega}{\sqrt{-\pi G\gamma^2\Delta\Lambda}}\ .
\end{equation}
In particular, in the low $\Lambda$ limit, the value $v_b$ approaches
\begin{equation}
    v_b \approx \frac{\omega}{\sqrt{3\pi G}}\ ,
\end{equation}
which corresponds to the bounce value of $v(\phi)$ of the LQC universe with scalar field momentum $p_{
\phi}=\hbar\omega$ and vanishing $\Lambda$.

\bibliography{main.bib}% Produces the bibliography via BibTeX.
\end{document}